\documentclass[aps,prc,twocolumn,superscriptaddress,nofootinbib]{revtex4}
\usepackage{graphicx}
\usepackage{amsmath}
\usepackage{amssymb}
\usepackage{hyperref}
\usepackage[usenames,dvipsnames]{xcolor}

\newcommand{\sNN}{s_\mathrm{NN}}

\begin{document}
\title{Estimation of the shear viscosity at finite net-baryon
  density from A+A collision data at $\sqrt{\sNN} = 7.7$--$200$ GeV}

\author{Iu.~A.~Karpenko}
\email{karpenko@fias.uni-frankfurt.de}
\affiliation{Frankfurt Institute for Advanced Studies,
  Ruth-Moufang-Stra{\ss}e 1, 60438 Frankfurt am Main, Germany}
\affiliation{Bogolyubov Institute for Theoretical Physics,
  Ul.~Metrolohichna, 14-b, 03680 Kiev, Ukraine}

\author{P.~Huovinen}
\affiliation{Frankfurt Institute for Advanced Studies,
  Ruth-Moufang-Stra{\ss}e 1, 60438 Frankfurt am Main, Germany}
\affiliation{Institute for Theoretical Physics, Goethe-Universit\"at,
  Max-von-Laue-Stra{\ss}e 1, 60438 Frankfurt am Main, Germany}

\author{H.~Petersen}
\affiliation{Frankfurt Institute for Advanced Studies,
  Ruth-Moufang-Stra{\ss}e 1, 60438 Frankfurt am Main, Germany}
\affiliation{Institute for Theoretical Physics, Goethe-Universit\"at,
  Max-von-Laue-Stra{\ss}e 1, 60438 Frankfurt am Main, Germany}
\affiliation{GSI Helmholtzzentrum für Schwerionenforschung GmbH,
  Planckstr. 1,  64291 Darmstadt, Germany}

\author{M.~Bleicher}
\affiliation{Frankfurt Institute for Advanced Studies,
  Ruth-Moufang-Stra{\ss}e 1, 60438 Frankfurt am Main, Germany}
\affiliation{Institute for Theoretical Physics, Goethe-Universit\"at,
  Max-von-Laue-Stra{\ss}e 1, 60438 Frankfurt am Main, Germany}

\begin{abstract}
  Hybrid approaches based on relativistic hydrodynamics and transport theory
  have been successfully applied for many years for the dynamical description of
  heavy ion collisions at ultrarelativistic energies. In this work a
  new viscous hybrid model employing the hadron transport approach
  UrQMD for the early and late non-equilibrium stages of the reaction,
  and 3+1 dimensional viscous hydrodynamics for the hot and dense
  quark-gluon plasma stage is introduced. This approach includes the
  equation of motion for finite baryon number, and employs an equation
  of state with finite net-baryon density to allow for calculations in a
  large range of beam energies. The parameter space of the model is
  explored, and constrained by comparison with the experimental data
  for bulk observables from SPS and the phase I beam energy scan at
  RHIC. The favored parameter values depend on energy, but allow to
  extract the effective value of the shear viscosity coefficient over
  entropy density ratio $\eta/s$ in the fluid phase for the whole energy region under investigation. The estimated value of $\eta/s$
  increases with decreasing collision energy, which may indicate that
  $\eta/s$ of the quark-gluon plasma depends on baryochemical potential
  $\mu_B$.
\end{abstract}

\maketitle

\section{Introduction}
Ultra-relativistic heavy ion collisions allow to investigate the
properties of strongly interacting matter under extreme conditions. At
high temperatures and/or high net-baryon densities a new state of
matter, the so-called quark-gluon plasma, QGP, is formed. The two
main goals of heavy ion research are the exploration of the phase
diagram of quantum chromodynamics and the determination of transport
coefficients of this new state of matter.

The studies of high energy heavy-ion collisions at the Large Hadron
Collider (LHC) at CERN and the Relativistic Heavy Ion Collider (RHIC)
at Brookhaven National Laboratory have
revealed that the quark-gluon plasma behaves like an almost perfect
fluid. In recent years, so-called hybrid approaches 
\cite{Hirano:2005xf,Nonaka:2005aj,Petersen:2008dd,Werner:2010aa,Song:2010aq} based on
(viscous) relativistic hydrodynamics for the hot and dense stage
coupled to hadron transport approaches for the decoupling stage of the
reaction have been applied with great success to extract average
values of the shear viscosity over entropy ratio $\eta/s$. The
results are very close to the conjectured universal limit of
$\eta/s=\frac{1}{4\pi}$, based on the anti--de Sitter/conformal field
theory (AdS/CFT) correspondence~\cite{Policastro:2001yc}. For example,
the values extracted in Ref.~\cite{Gale:2012rq} are $\eta/s = 0.12$ for
collisions at RHIC, and $\eta/s = 0.2$ at the LHC.

One expects the formation of partonic matter in heavy ion
collisions at ultra-relativistic energies (see, e.g.,
Ref.~\cite{Gyulassy:2004zy}).
However, it is unknown at what collision
energy the transition from hadronic to partonic matter sets in.
In addition, as the collisions at lower energies probe the phase diagram
at larger net-baryon densities, it may be possible to experimentally
see signs of the theoretically predicted critical
point~\cite{Stephanov:1999zu} and the first-order phase transition
beyond it. To investigate these questions the so-called beam energy scan
(BES) programs at SPS (NA49, NA61 experiments) and at RHIC (STAR, PHENIX experiments) were started. One of the surprises of the stage I of the 
BES program at RHIC has been that the $p_T$-differential elliptic flow,
$v_2(p_T)$, of charged hadrons does not change significantly when the
collision energy is reduced from $\sqrt{s_\mathrm{NN}} = 200$ to 
$\sim 20$ GeV~\cite{Adamczyk:2012ku}. The large values of elliptic
flow measured at $\sqrt{s_\mathrm{NN}} = 200$ GeV collisions were
taken as a sign of very low shear viscosity of the matter formed in
these collisions. Thus, the large $v_2(p_T)$ measured in collisions at
lower energy leads to the question how $\eta/s$ changes as function of
net-baryon density and baryochemical potential
$\mu_B$~\cite{Denicol:2013nua}.

Unfortunately, many of the hydrodynamical and hybrid models used to model
collisions at full RHIC and LHC energies are not directly applicable
to collisions at RHIC BES and CERN SPS energies, nor to collisions at even lower
energies in the future at Facility for Antiproton and Ion Research
(FAIR), Nuclotron-based Ion Collider Facility (NICA) and the stage II
of the BES program at RHIC. The simplifying approximations of boost
invariance and zero net-baryon density are not valid, and different
kinds of non-equilibrium effects play a larger role. To overcome these
limitations, a novel hybrid approach has been developed. This
approach is based on the Ultra-relativistic Quantum Molecular Dynamics
(UrQMD) transport~\cite{Bass:1998ca} for the non-equilibrium early and
late stages, and on a (3+1)-dimensional viscous hydrodynamical
model~\cite{Karpenko:2013wva} for the hot and dense stage of the
reaction.

In this paper, this approach is applied to extract the shear viscosity
coefficient over entropy density ratio of strongly interacting matter
from the heavy-ion collision data at RHIC beam energy scan
energies in the broad range $\sqrt{s_{\rm NN}}=7.7$--$200$ GeV. The
details of the model are explained in Section~\ref{model},
and the generic effects of finite shear viscosity on the hydrodynamical
expansion are described in Section~\ref{sensitivity}. The sensitivity
of particle yields and spectra to the parameters for the initial and
final state transitions is explored in Section~\ref{parameters}.
Section~\ref{results} contains the main results of this work including
the estimated values of the effective shear viscosity over entropy density 
ratio as a
function of beam energy. Finally, the main conclusions are
summarized and an outlook on future work is given in
Section~\ref{summary}.

\section{Model Description}
\label{model}

Our hybrid approach combines the UrQMD transport model~\cite{Bass:1998ca}
for the early and late stages of the evolution with a dissipative
hydrodynamical model, called \texttt{vHLLE}~\cite{Karpenko:2013wva}, for the hot and dense
stage. The distinguishing features of the present model are that the fluid
dynamical expansion is solved numerically in all three spatial
dimensions without assuming boost invariance nor cylindrical symmetry,
the equations of motion for finite net-baryon and charge densities are
explicitly included and, in contrast to the standard UrQMD hybrid
approach (UrQMD-3.4 at urqmd.org)~\cite{Petersen:2008dd,Auvinen:2013sba},
dissipation in the form of shear viscosity is included in the
hydrodynamical stage. Different to our previous
studies~\cite{Karpenko:2013ksa,Karpenko:2013ama}, event-by-event
fluctuations are now included. The hadronic cascade operates with the full phase-space distribution of the
final particles which allows for a proper comparison to experimental
data.

\subsection{Pre-thermal Phase}

The UrQMD string/hadronic cascade is used to describe the primary
 collisions of the nucleons, and to create the initial state of the
 hydrodynamical evolution. The two nuclei are initialized according to
 Woods-Saxon distributions and the initial binary interactions proceed
 via string or resonance excitations, the former process being dominant
 in ultrarelativistic collisions (including the BES collision
 energies). All the strings are fragmented into hadrons before the
 transition to fluid phase (fluidization) takes place, although not all
 hadrons are yet fully formed at that time, i.e., they do not yet have
 their free-particle scattering cross sections, and thus do not yet
 interact at all. The hadrons before conversion to fluid should not be
 considered physical hadrons, but rather marker particles to describe
 the flow of energy, momentum and conserved charges during the
 pre-equilibrium evolution of the system. The use of UrQMD to
 initialise the system allows us to describe some of the
 pre-equilibrium dynamics and dynamically generates event-by-event
 fluctuating initial states for hydrodynamical evolution.

The interactions in the pre-equilibrium UrQMD evolution are allowed
until a hypersurface of constant Bjorken proper time
$\tau_0=\sqrt{t^2-z^2}$ is reached, since the hydrodynamical code is
constructed using the Milne coordinates $(\tau,x,y,\eta)$, where $\tau
= \sqrt{t^2-z^2}$~\cite{Karpenko:2013wva}. The UrQMD evolution,
however, proceeds in Cartesian coordinates $(t,x,y,z)$, and thus
evolving the particle distributions to constant $\tau$ means evolving
the system until large enough time $t_{l}$ in such a way that the
collisional processes and decays are only allowed in the domain
$\sqrt{t^2-z^2} < \tau_0$. The resulting particles on $t = t_l$
surface are then propagated backwards in time to the $\tau = \tau_0$
surface along straight trajectories to obtain an initial state for the
hydrodynamic evolution.

\begin{figure}
\includegraphics[width=0.45\textwidth]{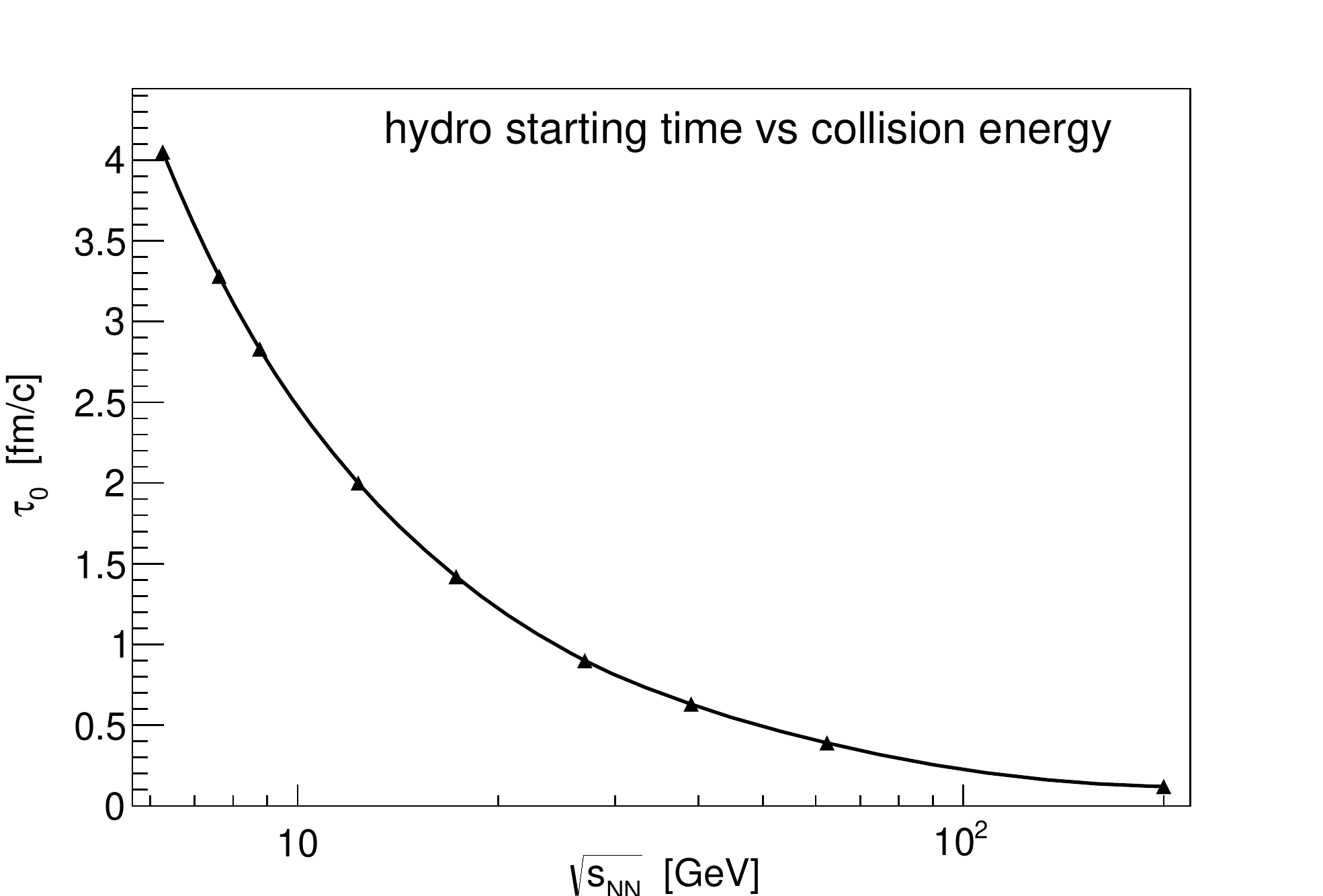}\\
\caption{\label{figStartTime} The earliest possible starting time of the hydrodynamic
  evolution as a function of $\sqrt{s_{NN}}$ according to
  Eq.~\ref{eqTau0}.}
\end{figure}

The lower limit for the starting time of the hydrodynamic evolution depends on 
the collision energy according to
\begin{equation}
\tau_0=2R/\sqrt{(\sqrt{\sNN}/2m_N)^2-1}, \label{eqTau0}
\end{equation}
which corresponds to the average time, when two nuclei have passed
through each other, i.e., all primary nucleon-nucleon
collisions have happened. This is the earliest possible moment in
time, where approximate local equilibrium can be assumed. The
$\tau_0$ values calculated according to this formula are plotted in
Fig.~\ref{figStartTime}.

To perform event-by-event hydrodynamics using fluctuating initial
conditions every individual UrQMD event is converted to an initial state
profile. As mentioned, the hadron transport does not lead to an
initial state in full local equilibrium, and the thermalization of the
system at $\tau=\tau_0$ has to be artificially enforced.  The energy and
momentum of each UrQMD particle at $\tau_0$ is distributed to the
hydrodynamic cells $ijk$ assuming Gaussian density
profiles
\begin{align} \label{Gauss1}
\Delta P^\alpha_{ijk} & = P^\alpha \cdot C\cdot\exp\left(-\frac{\Delta x_i^2+\Delta y_j^2}{R_\perp^2}-\frac{\Delta\eta_k^2}{R_\eta^2}\gamma_\eta^2 \tau_0^2\right) \\
\Delta N^0_{ijk}&=N^0 \cdot C\cdot\exp\left(-\frac{\Delta x_i^2+\Delta y_j^2}{R_\perp^2}-\frac{\Delta\eta_k^2}{R_\eta^2}\gamma_\eta^2 \tau_0^2\right), \label{Gauss2}
\end{align}
where $\Delta x_i$, $\Delta y_j$, $\Delta \eta_k$ are the differences
between particle's position and the coordinates of the hydrodynamic
cell $\{i,j,k\}$, and $\gamma_\eta={\rm cosh}(y_p-\eta)$ is the
longitudinal Lorentz factor of the particle as seen in a frame moving
with the rapidity $\eta$. The normalization constant $C$ is calculated
from the condition that the discrete sum of the values of the Gaussian
in all neighboring cells equals one. The resulting $\Delta P^\alpha$
and $\Delta N^0$ are transformed into Milne coordinates and added
to the energy, momentum and baryon number in each cell. This procedure
ensures that in the initial transition from transport to hydrodynamics
energy, momentum and baryon number are conserved.

For the present study energy and momentum of the initial particles are
converted at $\tau_0$ into a perfectly equilibrated fluid, i.e., the
initial values for the viscous terms in the energy-momentum tensor are
set to zero: $\pi^{\mu\nu}(\tau_0)=\Pi(\tau_0)=0$. In other words the
$T^{0\mu}$ components of the energy-momentum tensor stay the same, but
the $T^{ij}$ components change, when we switch from UrQMD to the
fluid. Thus, we do not consider how much the energy-momentum tensor of
UrQMD deviates from the ideal fluid energy-momentum tensor, but leave
this topic for further studies.

\begin{figure}
\includegraphics[width=0.5\textwidth]{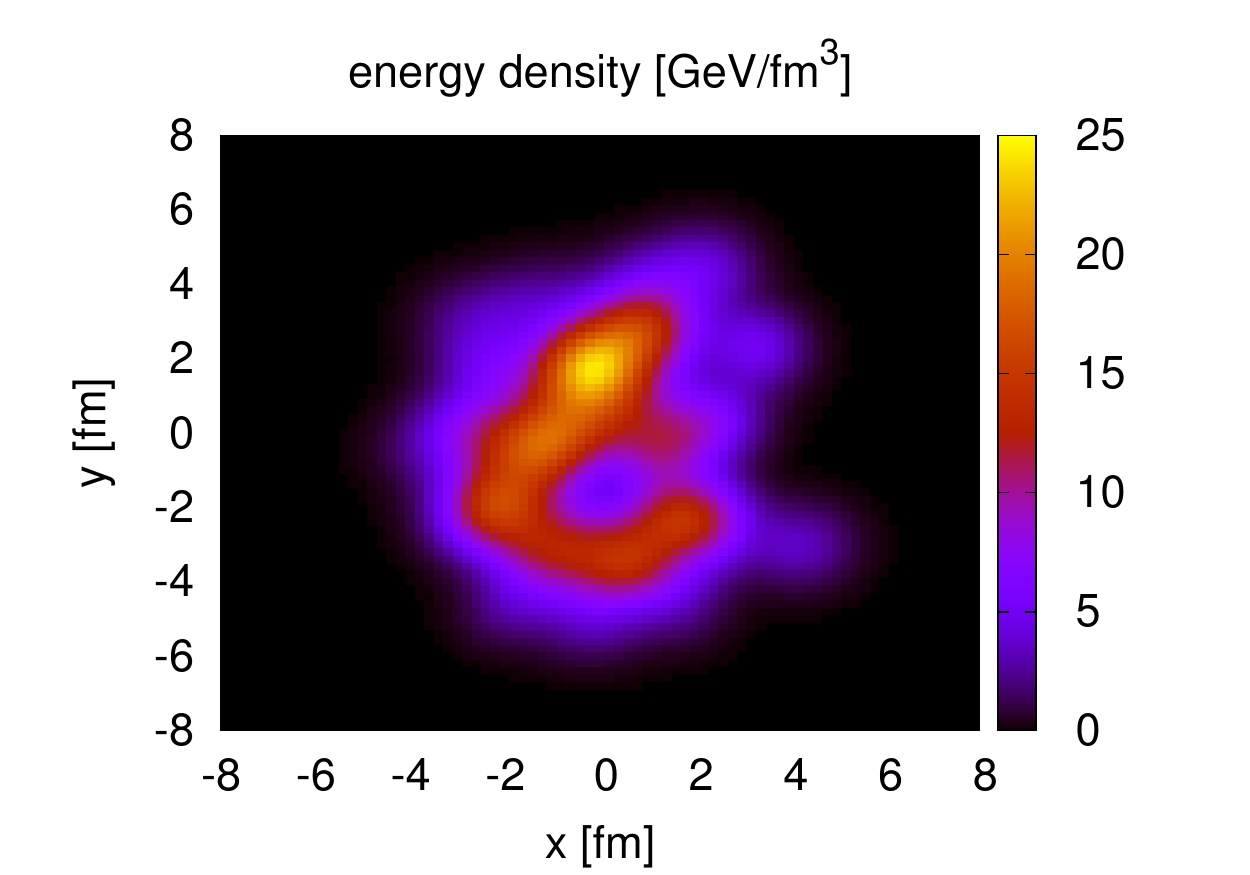}
\caption{An example of a fluctuating (single event) initial energy density profile in
  the transverse plane at $\eta=0$. The profile is obtained with
  $R_\perp=R_\eta=1$~fm Gaussian smearing and corresponds to a 20-30\%
  Au-Au collision at $\sqrt{\sNN}=39$~GeV.}\label{fig-ic-ebe}
\end{figure}

One typical example of the initial energy density distributions in the
transverse plane at midrapidity for one event is presented in
Fig.~\ref{fig-ic-ebe}. The parameters $R_\perp$ and $R_\eta$ regulate
the granularity of the initial state. At the same time they influence
the initial entropy of the hydrodynamic evolution, while the total
initial energy and momentum are always fixed to be equal to the energy
and momentum of the pre-equilibrium UrQMD event. The dependence of the
final results on these two parameters is discussed later in
Section~\ref{parameters}.

\subsection{Hydrodynamic Evolution} 

The (3+1)-dimensional viscous hydrodynamical code \texttt{vHLLE} is
described in full detail in Ref.~\cite{Karpenko:2013wva}. We
repeat here only its main features. The code solves the usual local
energy-momentum conservation equations
\begin{align}
 \partial_{;\nu}T^{\mu\nu}&=0, \label{hydro1}\\
 \partial_{;\nu}N_{\rm B,Q}^\nu&=0, \label{hydro2}
\end{align}
where $N^\nu_{\rm B}$ and $N^\nu_{\rm Q}$ are net baryon and electric charge
currents respectively, and the semicolon denotes the covariant
derivative. The calculation\footnote{Typical grid spacing used in the calculations: $\Delta x=\Delta y=0.2$~fm, $\Delta\eta=0.05-0.15$ and timestep $\Delta\tau=0.05-0.1$~fm/c depending on the collision energy. A finer grid with $\Delta x=\Delta y=0.125$~fm was taken to simulate peripheral collisions.}
is done in Milne coordinates $(\tau,x,y,\eta)$, where $\tau = \sqrt{t^2-z^2}$ and
$\eta=1/2\ln[(t+z)/(t-z)]$.

In the Israel-Stewart framework of causal dissipative
hydrodynamics~\cite{Israel-Stewart}, the dissipative currents are
independent variables. For the purpose of the present work we set the
bulk viscosity to zero, $\zeta/s=0$. We work in the Landau frame, where
the energy diffusion flow is zero, and neglect the baryon and charge
diffusion currents, which is equivalent to zero heat conductivity. For
the shear stress evolution we choose the relaxation time 
$\tau_\pi = 5\eta/(Ts)$, the coefficient $\delta_{\pi\pi} = 4/3 \tau_\pi$, and
approximate all the other
coefficients~\cite{Denicol:2012cn,Molnar:2014zha} by zero. For the
shear-stress tensor $\pi^{\mu\nu}$ we obtain the evolution equation
\begin{equation}
\left<u^\gamma \partial_{;\gamma} \pi^{\mu\nu}\right>
 =-\frac{\pi^{\mu\nu}-\pi_\text{NS}^{\mu\nu}}{\tau_\pi}
  -\frac 4 3 \pi^{\mu\nu}\partial_{;\gamma}u^\gamma, \label{evolutionShear}
\end{equation}
where the brackets denote the traceless and orthogonal to $u^\mu$ part
of the tensor and $\pi_\text{NS}^{\mu\nu}$ is the Navier-Stokes value
of the shear-stress tensor.

Another necessary ingredient for the hydrodynamic stage is the
equation of state (EoS) of the medium. We use the chiral
model EoS \cite{Steinheimer:2010ib}, which features correct asymptotic
degrees of freedom, i.e., quarks and gluons in the high temperature and
hadrons in the low-temperature limits, crossover-type transition
between confined and deconfined matter for all values of $\mu_B$ and
qualitatively agrees with lattice QCD data at $\mu_B=0$.

The tests to confirm the accuracy of the code have been reported in
Ref.~\cite{Karpenko:2013wva}. In particular the solutions have been
compared to the ideal Gubser solution~\cite{Gubser:2010ze} and to a
numerical solution of dissipative hydrodynamics calculated using the
\texttt{VISH2+1} hydro code~\cite{Song:2007ux}.

\subsection{Particlization and Hadronic Rescattering} 

It is well known that hydrodynamics loses its validity when the system
becomes dilute. To deal with this problem we apply the conventional
Cooper-Frye prescription \cite{Cooper:1974mv} to particlize the system
(convert the fluid to individual particles) at a hypersurface of
constant local rest frame energy density, and use the UrQMD cascade to
describe the further evolution of these particles. This switching
hypersurface is evaluated during the hydrodynamic evolution using the
Cornelius routine~\cite{Huovinen:2012is}, and as a default value for
the switching density we use $\epsilon_{\rm sw}=0.5$~GeV/fm$^3$, which
in the chiral model EoS corresponds to $T\approx 175$~MeV at $\mu_{\rm B}=0$. 
At this energy density the crossover transition is firmly on the
hadronic side, but the density is still a little higher than the
chemical freeze-out energy density suggested by the thermal
models~\cite{Becattini:2005xt}. Thus the hadronic transport can take
care of both chemical and kinetic decoupling of hadrons. We discuss
the sensitivity of the results to the value of the switching density
in section~\ref{parameters}.

As given by the Cooper-Frye prescription, the hadron distribution on
each point of the hypersurface is
\begin{equation}
p^0 \frac{d^3N_i(x)}{d^3p} = d\sigma_\mu p^\mu f(p\cdot u(x),T(x),\mu_i(x)).
 \label{CFp}
\end{equation}
The phase space distribution function $f$ is usually assumed to be the
one corresponding to a noninteracting hadron resonance gas in or close
to the local thermal equilibrium. This is inconsistent with mean
fields included in the chiral model EoS used during the evolution. To
solve this inconsistency we evaluate the switching surface using the
chiral model EoS, but use a free hadron resonance gas EoS to
recalculate the energy density, pressure, flow velocity $u^\mu$,
temperature, and chemical potentials from the ideal part of the
energy-momentum tensor and charge currents, and use these values to
evaluate the particle distributions on the switching surface. For
example the above mentioned temperature of $T\approx 175$ MeV in
chiral model EoS at zero baryon density and $\epsilon_{\rm sw}=0.5$~GeV/fm$^3$,
drops to $T\approx 165$ MeV in the free hadron resonance gas EoS.
This procedure ensures that
the total energy of the produced particles is reasonably close to the
overall energy flow through the particlization hypersurface (up to
negative contributions to the Cooper-Frye formula), although a small
error arises since we use a different energy density to evaluate the
position of the surface, and the properties of the fluid on
it\footnote{The exact procedure suggested in Ref.~\cite{Cheng:2010mm}
  requires a numerical solution of a cubic equation for each surface
  element, and is therefore too slow for event-by-event studies.}.
We have checked that in a case of event-averaged initialization, this
error is on the level of few percents. In addition, the conservation of energy and
momentum in the 3+1 dimensional numerical solution of
the fluid-dynamical equations using Milne coordinates is slightly violated as discussed in
Refs.~\cite{Karpenko:2013wva,Molnar:2014zha}.

To take into account the dissipative corrections to the distribution
function $f$, we use the well-known Grad's 14-moment ansatz for a
single component system, and assume that the correction is the same
for all hadron species. We evaluate the particle distribution in the rest
frame of the fluid at each surface element using the Cooper-Frye formula
\begin{multline}
\frac{d^3 \Delta N_i}{dp^* d({\rm cos}\,\theta)d\phi}=\underbrace{\frac{\Delta\sigma^*_\mu p^{*\mu}}{p^{*0}}}_{W_\text{residual}} 
\underbrace{p^{*2} f_\text{eq}(p^{*0};T,\mu_i)}_\text{isotropic} \\
\times\underbrace{\left[ 1+(1\mp f_\text{eq})\frac{p^*_\mu p^*_\nu \pi^{*\mu\nu}}{2T^2(\epsilon+p)} \right]}_{W_\text{visc}}. \label{DF-LRF-visc}
\end{multline}

The distribution function in Eq. (\ref{DF-LRF-visc}) is expressed in
terms of temperature and chemical potential(s), which implies a grand
canonical ensemble. This allows to do the particle sampling
independently on each surface element. To create an ensemble for
particles, we perform the following steps at each element
$\Delta\sigma_i$:
\begin{itemize}
\item First, the average number of hadrons of every sort is
  calculated:
$$\Delta N_i=\Delta\sigma_\mu u^\mu n_\text{i,th}=\Delta\sigma^*_0 n_\text{i,th}$$
\item For a given $\langle N_\text{tot}\rangle = \sum_i N_i$, the
  number of particles to be created is generated according to a
  Poisson distribution with a mean value $\langle N_\text{tot}\rangle$.
\item For each created particle, the type is randomly chosen based on
  the probabilities $N_i/N_\text{tot}$.
\item A momentum is assigned to the particle in two steps:
\begin{enumerate}
 \item The modulus of the momentum is sampled in the local rest frame
   of the fluid, according to the isotropic part of
   Eq.~(\ref{DF-LRF-visc}), and the direction of momentum is picked
   randomly in $4\pi$ solid angle.
 \item The correction for $W_\text{residual}$ or
   $W_\text{residual}\cdot W_\text{visc}$ in Eq.~(\ref{DF-LRF-visc}) is
   applied via rejection sampling: A random number $x$ in the range
   $[0,W_\text{max}]$ is generated. If $x<W$, the generated momentum
   is accepted, if not, the momentum generating procedure is repeated.
\end{enumerate}
\item The particle momentum is Lorentz boosted to the center
   of mass frame of the system.
\item The particle position is taken to be equal to the coordinate of the centroid of the corresponding surface element, except for the spacetime rapidity of the particle, which is uniformly distributed within the longitudinal size of the volume element.
\end{itemize}
For the current study, no correction over the grand canonical
procedure is made to effectively account for the exact conservation of the total baryon/electric charge,
strangeness and total energy/momentum for every sampled
event\footnote{For a suggested procedure to impose the conservation
  laws, see Ref.~\cite{Huovinen:2012is}.}. As a result, these
quantities fluctuate from event to event.

The generated hadrons are then fed into the UrQMD cascade. Since the
cascade accepts only a list of particles at an equal Cartesian time as
an input, the created particles are propagated backwards in time to
the time when the first particle was created. The particles are not allowed 
to interact in the cascade until their trajectories cross the particlization
hypersurface.

\section{Sensitivity to Shear Viscosity}
\label{sensitivity}

\begin{figure}
\includegraphics[width=0.47\textwidth]{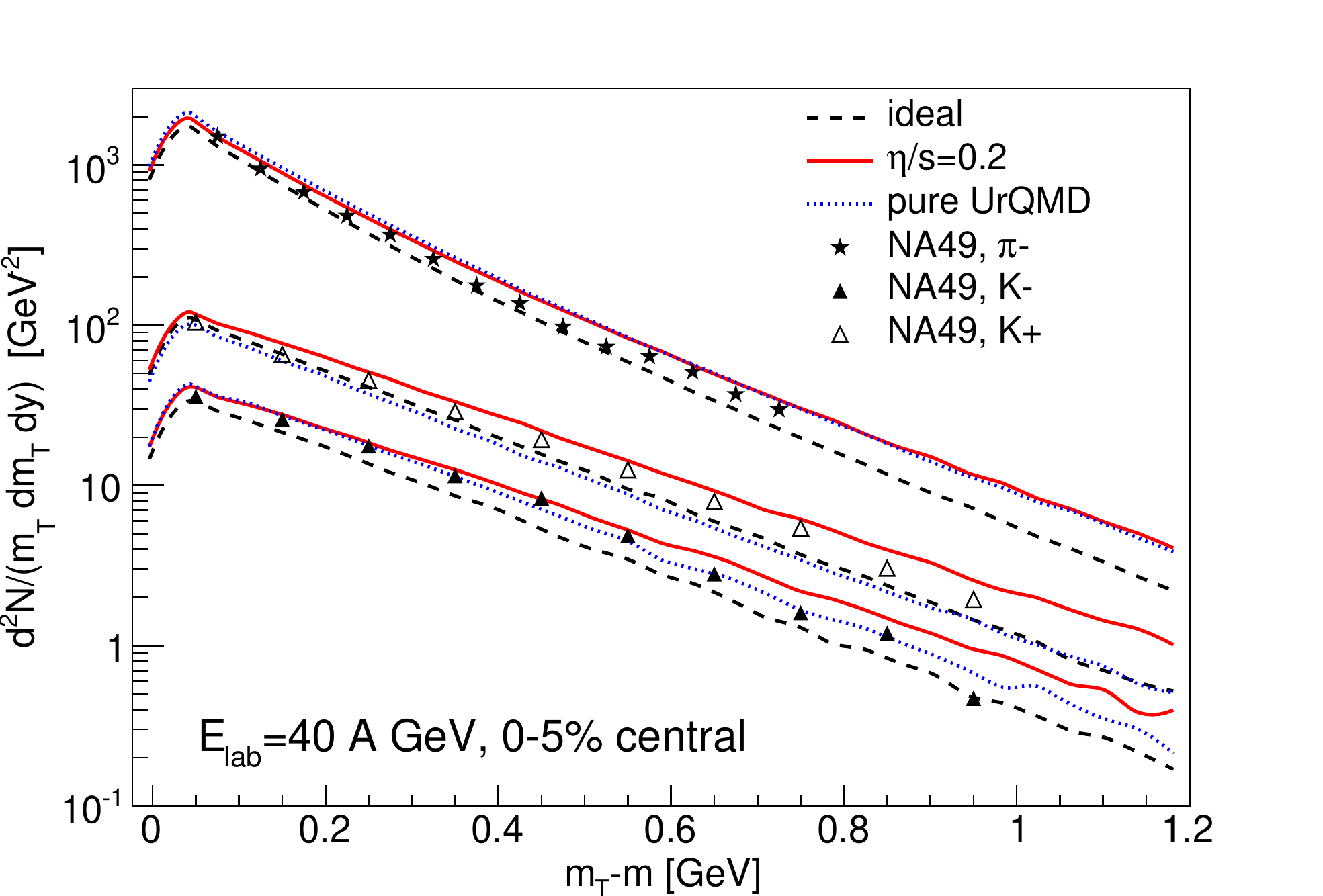}
\caption{(Color online) Transverse momentum spectra of negative pions, positive 
and negative kaons in $E_{\rm lab}=40$~A~GeV ($\sqrt{\sNN} = 8.8$ GeV)
  central Pb-Pb collisions. The experimental data from the NA49
  collaboration~\cite{Afanasiev:2002mx} are compared to the hybrid model
  calculations with $\eta/s=0$ (dashed lines) and $\eta/s=0.2$ (solid
  lines) in the hydrodynamic phase. The results from UrQMD model with
  no intermediate hydro phase (dubbed as ``pure UrQMD'') are shown
  with dotted lines.}\label{figpt}
\end{figure}

The overall effects of shear viscosity on hydrodynamical expansion
have been extensively discussed in the
literature~\cite{Song:2007ux,Muronga:2001zk,Baier:2006gy,Teaney:2003kp}.
Here we show that the results from high energy collisions, e.g.,
entropy increase, enhancement of transverse and inhibition of
longitudinal expansion, and suppression of anisotropies are also manifested
in calculations at lower collision energies.

We have carried out event-by-event simulations for different collision
energies, centralities, and two fixed values of shear viscosity:
$\eta/s=0$ (ideal hydro evolution) and $\eta/s=0.2$. For these
simulations we use the values of the Gaussian radii for the particles'
energy/momentum deposition $R_\perp=R_\eta=1$~fm (see
Eqs.~(\ref{Gauss1}) and~(\ref{Gauss2})). The initial time is chosen
according to Eq.~(\ref{eqTau0}), however for the collisions at
energies equal or higher than $\sqrt{s_\mathrm{NN}} = 27$~GeV we set
$\tau_0 = 1$ fm/$c$.

To reduce the need for CPU time, we use so called oversampling
technique, as in Ref.~\cite{Holopainen:2010gz}. For each collision
energy, centrality, and parameter set we have created around 500
hydrodynamic events with randomly generated initial conditions. For
each hydrodynamic event, or transition hypersurface, we generate
$N_{\rm oversample}=50-100$ final state events, which results in
$25000-50000$ events used to calculate observables. We have checked
that the oversampling procedure does not significantly affect the
final observables by creating 1000 or 10000 hydrodynamic events, with
$N_{\rm oversample}=20\ {\rm and}\ 2$ respectively, for several
parameter sets. In both cases the calculated observables agreed within
statistical errors.

\begin{figure}
\includegraphics[width=0.47\textwidth]{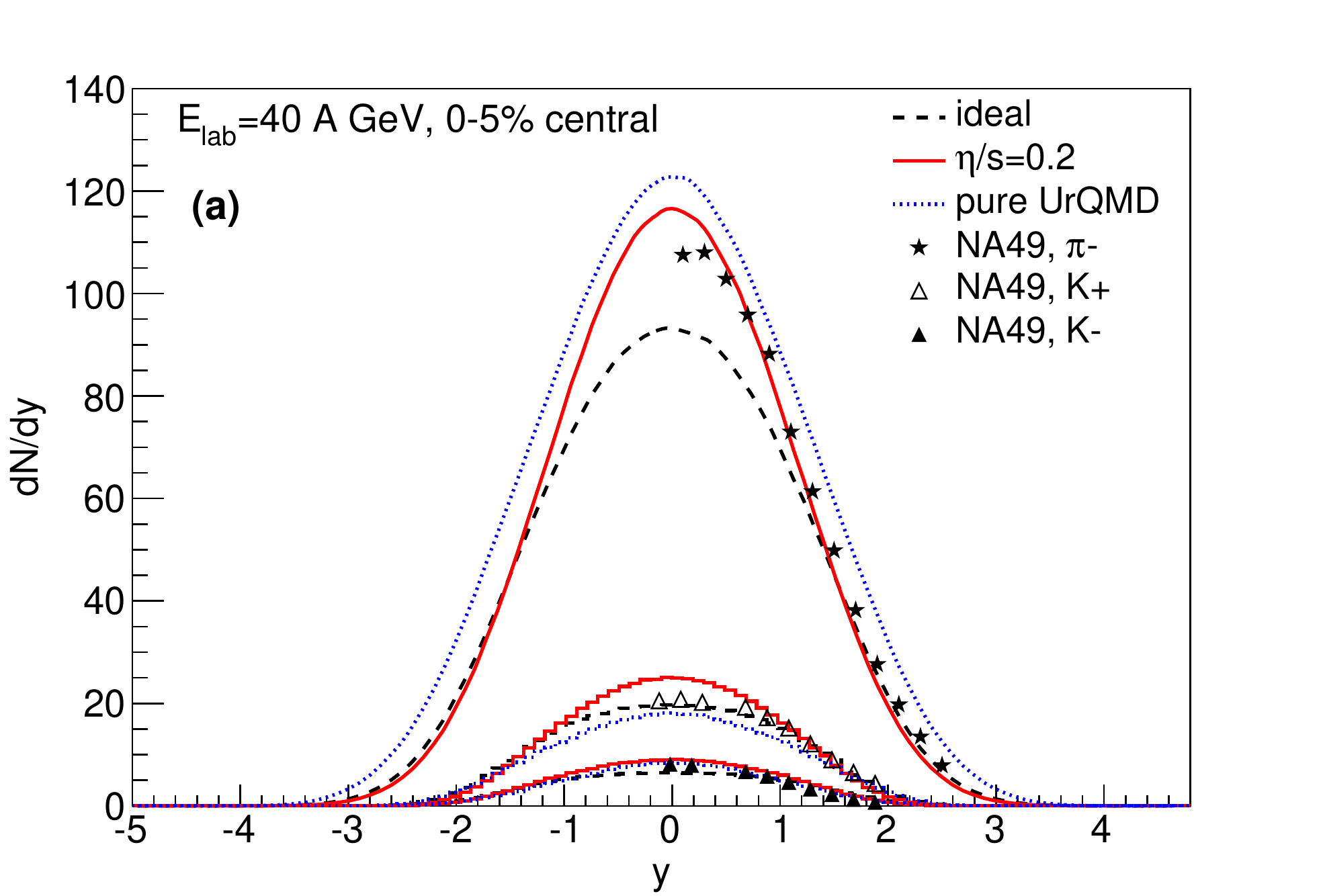}\\
\includegraphics[width=0.47\textwidth]{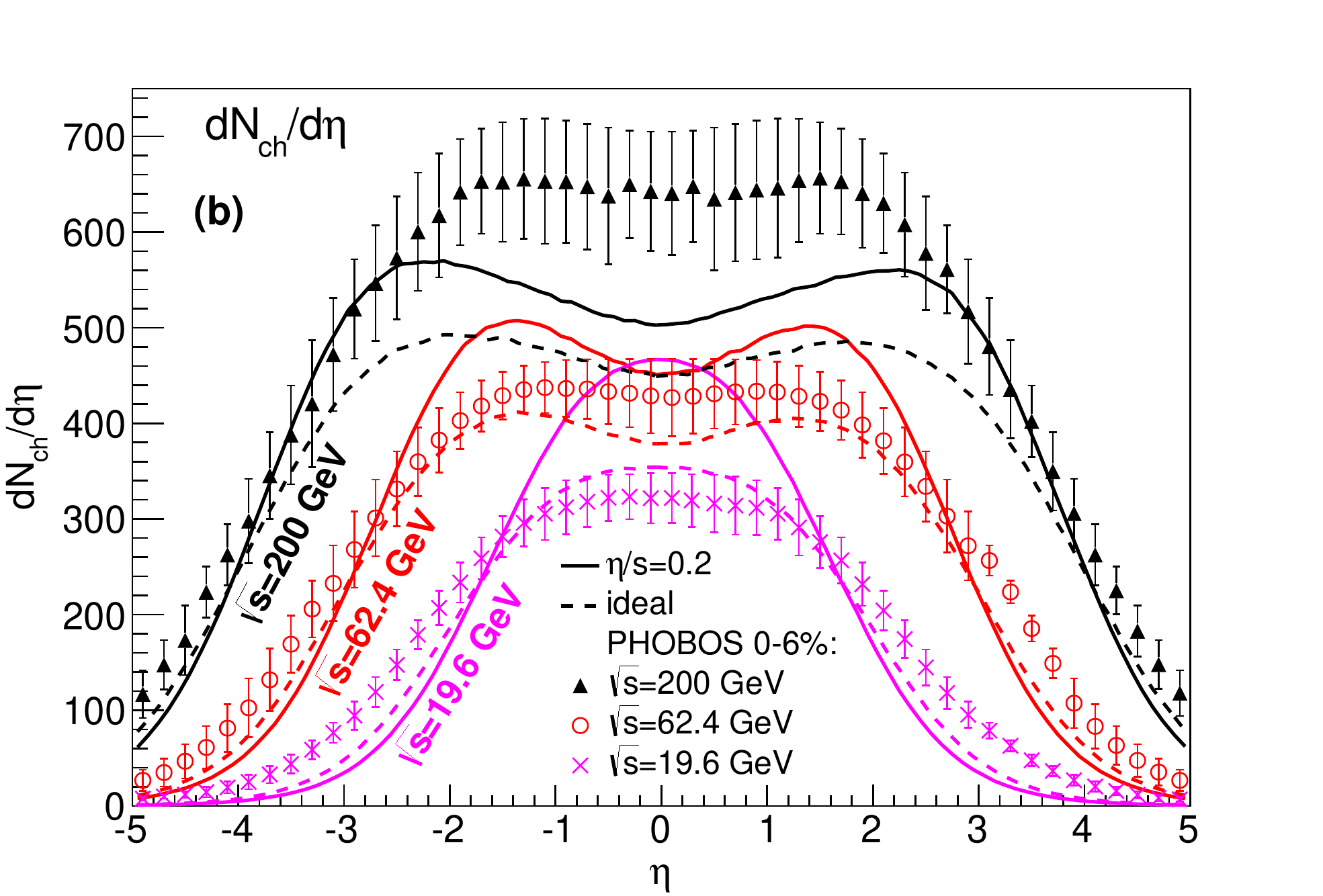}
\caption{(Color online) Pion and kaon $dN/dy$ in $E_{\rm lab}=40$~A~GeV
  ($\sqrt{\sNN}=8.8$ GeV) central Pb-Pb collisions (top) and
  charged hadron $dN/d\eta$ distributions at $\sqrt{\sNN}=19.6$, 39, 62.4 and
  200~GeV central Au-Au collisions (bottom). The experimental data
  from the NA49~\cite{Afanasiev:2002mx} and the PHOBOS
  collaborations~\cite{Alver:2010ck} are compared to the hybrid model
  calculations with $\eta/s=0$ (dashed lines) and $\eta/s=0.2$ (solid
  lines) in the hydrodynamic phase.}\label{figdNdeta}
\end{figure}

The available experimental data set for the basic bulk hadron
observables at the BES energies is inhomogeneous. (Pseudo)rapidity
spectra of all charged hadrons for Au-Au collisions are available from
the PHOBOS analysis \cite{Alver:2010ck} for $\sqrt{\sNN}=19.6$, 62.4 and
200~GeV energies only. The $p_T$ spectra are published for
$\sqrt{\sNN}=62.4$~GeV by the PHOBOS collaboration~\cite{Back:2006tt} and
for $\sqrt{\sNN}=200$~GeV by the PHENIX collaboration~\cite{Adler:2003cb}.
To cover the lower collision energies we use $dN/dy$ and $p_T$-spectra
from the NA49~\cite{Afanasiev:2002mx} collaboration for Pb-Pb collisions at $E_{\rm lab}=40$
and $158$~A~GeV, which correspond to $\sqrt{\sNN}=8.8$ and 17.6~GeV, and
set up the simulations accordingly for Pb-Pb system. For the elliptic
flow we compare to the STAR results at $\sqrt{\sNN}=7.7$, 11.5, 19.6, 27, 39 GeV \cite{Adamczyk:2012ku} and 200 GeV \cite{Adams:2004bi} collision energies. In the model we define the centrality classes as impact parameter intervals based on the optical Glauber model estimates \cite{Eskola:1988yh,Miskowiec}.

\begin{figure}
\includegraphics[width=0.47\textwidth]{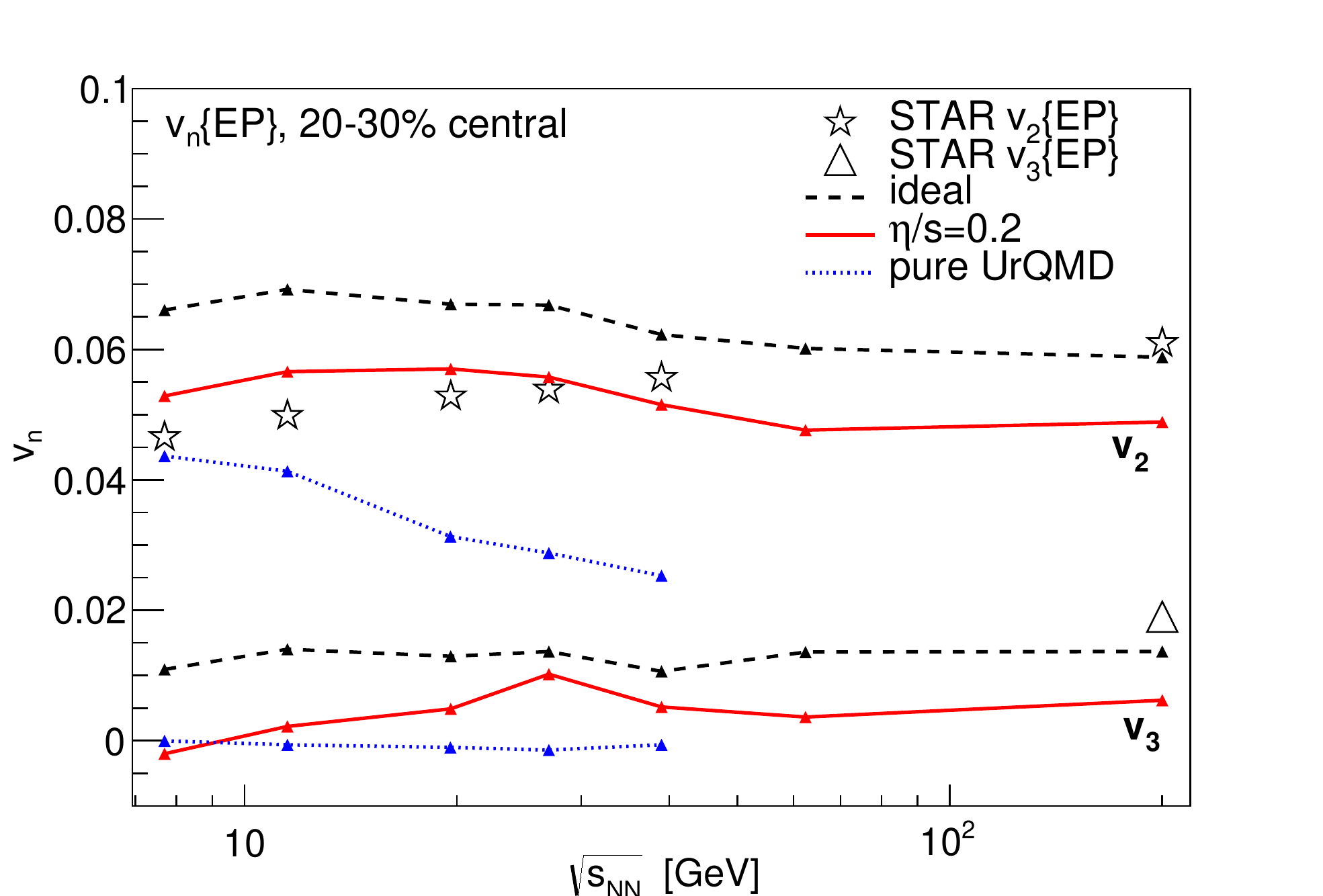}
\caption{(Color online) $p_T$ integrated ($0.2<p_T<2.0$~GeV and $|\eta|<1$) 
elliptic ($v_2$) and triangular ($v_3$)
  flows of all charged hadrons in 20-30\% central Au-Au collisions as
  a function of collision energy, calculated with the event-plane
  method. The elliptic and triangular flow data is from the STAR
  collaboration~\cite{Adamczyk:2012ku,Adamczyk:2013waa}. The solid line depicts the
  calculation with $\eta/s = 0.2$, the dashed line with $\eta/s = 0$
  whereas the dotted line corresponds to the ``pure'' UrQMD
  calculation with no intermediate hydrodynamic stage.}\label{figv2}
\end{figure}

The transverse momentum distributions of identified particles at
$\sqrt{s_\mathrm{NN}} = 8.8$ GeV ($E_{\rm lab}=40$~A~GeV) collision,
and (pseudo)rapidity distributions of identified or charged particles
at collision energies $\sqrt{s_\mathrm{NN}} = 8.8$--200 GeV are shown
in Figs.~\ref{figpt} and~\ref{figdNdeta}, respectively. As can be
seen, the inclusion of shear viscosity in the hydrodynamic phase
hardens the $p_T$ spectra, and increases $dN/dy$ (and similarly
$dN/d\eta$) at midrapidity, squeezing the overall rapidity
distribution. This effect can be attributed to the effect of shear
viscosity on the strong longitudinal expansion of the system in the
initial state for the hydrodynamic phase. Shear viscosity attempts to
isotropize the expansion by decelerating it in
the longitudinal direction and accelerating it in the transverse
direction. The energy of the hydrodynamic system is always conserved,
whereas additional entropy is produced during the viscous hydrodynamic
evolution, which explains the increased total particle
multiplicity. Comparing to the experimental data one observes that
$\eta/s=0.2$ gives a good estimate of the rapidity and transverse momentum
distributions at the lowest collision energy point
$\sqrt{s_\mathrm{NN}} = 8.8$ GeV ($E_{\rm lab}=40$~A~GeV), while
overestimating $dN/d\eta$ at midrapidity for the rest of collision
energies except for the highest energy, $\sqrt{\sNN}=200$~GeV, where
we underestimate the PHOBOS results.

In Fig.~\ref{figv2} the $p_T$-averaged elliptic and triangular flow
coefficients $v_2$ and $v_3$ are shown as a function of collision
energy. The flow coefficients are calculated using the event-plane
method as described in Ref.~\cite{Holopainen:2010gz}, including the
event plane resolution correction. As expected, the elliptic and
triangular flow coefficients are suppressed by the shear
viscosity. However, comparing the results for $\eta/s=0.2$ to the STAR
experimental results at $20-30\%$ centrality we find that the
suppression is too weak for $\sqrt{\sNN}\lesssim 30$~GeV and too
strong otherwise. The latter is consistent with the fact that the
optimal value of $\eta/s$ required to fit the elliptic flow data at
$\sqrt{s_\mathrm{NN}}=200$~A~GeV is $\eta/s=0.08$ assuming the initial
energy density profile from Monte Carlo Glauber approach
\cite{Song:2010mg}. Another particular feature of both $v_2$ curves is
that, in the region $\sqrt{s_\mathrm{NN}}\approx20$--62~GeV the
elliptic flow decreases as a function of $\sqrt{\sNN}$. If we do not
limit the initial time $\tau_0$ from below at energies
$\sqrt{s_\mathrm{NN}} > 25$ GeV, but take it directly from
Eq.~(\ref{eqTau0}), we do not see this decrease, but $v_2$ increases
monotonously with increasing collision energy. Thus we expect that the
reason for the nonmonotonous behavior is in our choice for the initial
time of the hydrodynamic evolution.

The results from the standard UrQMD cascade (without intermediate
hydrodynamic phase) are also shown for comparison on Figs.~\ref{figpt}
and~\ref{figdNdeta} with dotted lines. One may conclude that, whereas
standard UrQMD does a good job for $p_T$-spectra and rapidity
distributions at the lowest energy, it clearly underestimates
$v_2$ when the collision energy increases 
(which repeats the conclusion about the $v_2$ excitation function from
Ref.~\cite{Petersen:2009vx}, and later results from $v_3$ analysis in
Ref.~\cite{Auvinen:2013sba}). This is an indication of too large
viscosity of the high-density medium and served historically
as a motivation to introduce the intermediate hydrodynamic stage.

\section{Investigation of Parameter space}
\label{parameters}

\begin{figure}
\includegraphics[width=0.52\textwidth]{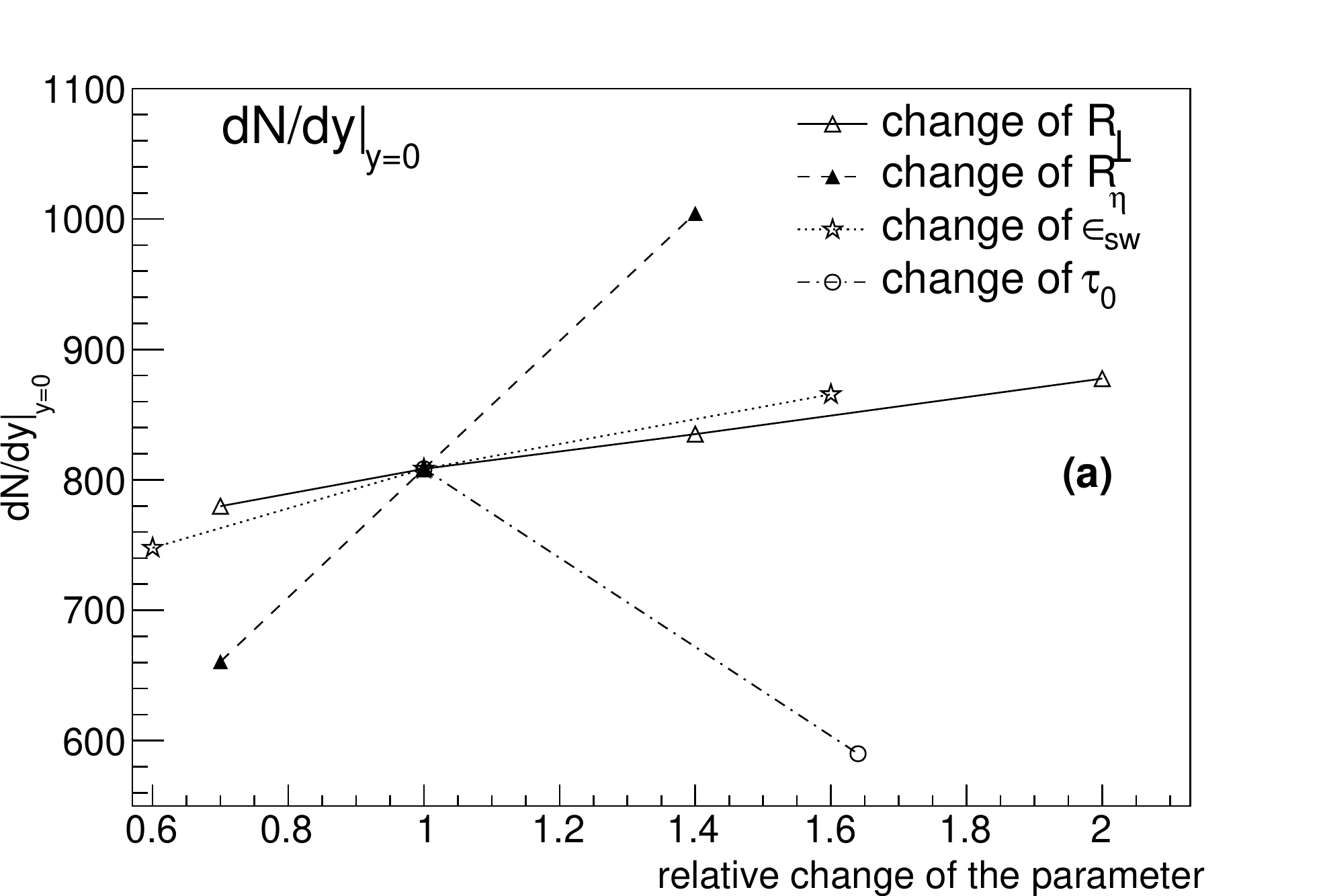}\\
\includegraphics[width=0.52\textwidth]{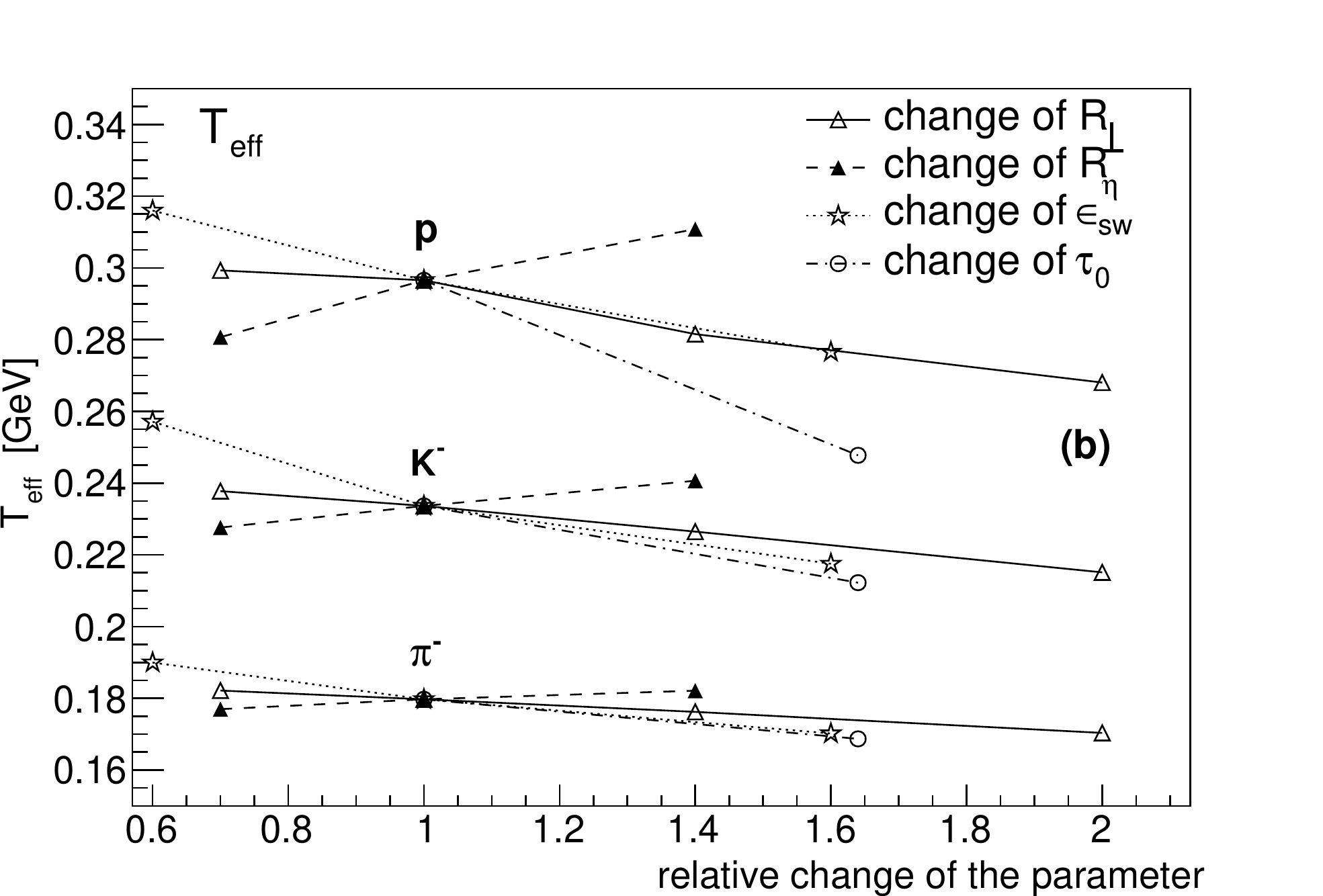}
\caption{Parameter dependence of the total yield at midrapidity (top)
  and the effective temperature of pion, kaon and proton
  $p_T$ spectra (bottom) in 0-5\% central
  Au-Au collisions at $\sqrt{\sNN}=19.6$~GeV.}\label{deps1}
\end{figure}

After investigating the generic influence of a finite shear viscosity
during the hydrodynamic evolution on basic bulk
observables, it is clear that we cannot fit all the available
experimental data using the same set of parameters\footnote{
 The internal parameters of UrQMD, e.g., particle properties and cross
 sections, are fixed by experimental data as explained in
 Ref.~\cite{Petersen:2008kb}. The effects of changes in resonance
 properties were studied in Ref.~\cite{Gerhard:2012fj}. It was found that
 if the changes stay within experimentally acceptable limits, the effects
 on final particle distributions are small.}. Thus we have to
adjust the model parameters according to the collision energy before
drawing any conclusions about the physical properties of the system.

\begin{figure}
\includegraphics[width=0.47\textwidth]{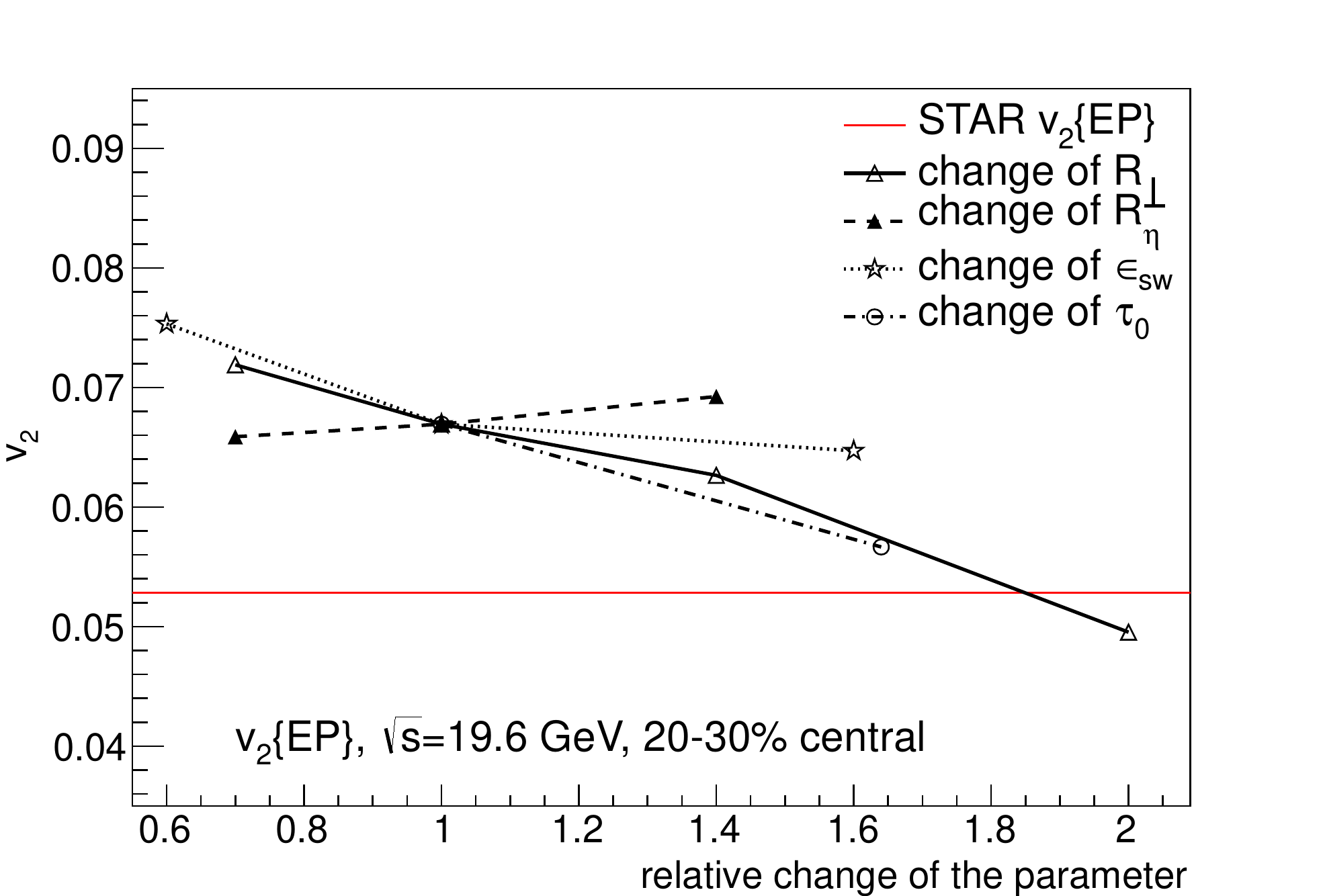}
\caption{(Color online) Parameter dependence of $p_T$ integrated elliptic flow 
$v_2$ of charged hadrons in 20-30\% central Au-Au collisions at
  $\sqrt{\sNN}=19.6$~GeV. The experimental value of the elliptic flow is
  shown with a solid red line for comparison.}\label{deps2}
\end{figure}

In this section we study systematically the sensitivities of the
particle yield at midrapidity, which is a measure for the final
entropy, the effective slope parameter that measures the strength of
the transverse expansion, and the anisotropic flow to the main
parameters of the model. Due to the limited space, and to emphasize
the main features of the dependencies, we restrict ourselves to one
collision energy, $\sqrt{\sNN}=19.6$~GeV, in the middle of the
investigated range. Since the influence of shear viscosity was
discussed above, we now concentrate on the remaining parameters of the
model: the two Gaussian radii $R_\perp$ and $R_\eta$ for the initial
distribution of energy, momentum and charges, the starting time for
the hydro phase $\tau_0$, and the energy density $\epsilon_{\rm sw}$
when the switch to the hadronic cascade happens. The default case is
$R_{\perp}=R_\eta=1.0$~fm, $\tau_0=1.22$~fm/c (calculated according to
Eq.~\ref{eqTau0}), $\eta/s=0$ (for simplicity) and 
$\epsilon_{\rm sw}=0.5$~GeV/fm$^3$. The dependencies are presented in
Figs.~\ref{deps1} and~\ref{deps2}, where each curve corresponds to the
variation of only one of the parameters, while keeping the default
values for the others. All values are normalized to their default
values to allow a direct comparison to each other. The effective
temperatures of the hadron spectra in the lower panel of
Fig.~\ref{deps1} are defined as the parameter of the exponential fit:
$$\frac{dN}{m_T dm_T dy}=C\exp\left(-\frac{m_T}{T_{\rm eff}}\right),$$
where the $m_T-m$ range is [0.2--1]~GeV for pions and protons and [0.05--1]~GeV for kaons\footnote{Smaller $m_T-m$ range for pions and protons is taken since the lowest $m_T-m$ part of the spectrum has a different slope than the intermediate $m_T-m$ range.}.
In general we do observe only a very weak dependence on the parameters,
that is less than 10\% for a 10\% change in parameters.
 The observed dependencies can be summarized as:

\begin{itemize}
 \item Increased $R_\perp$ smoothens the initial energy density
   profile in the transverse plane, which leads to smaller gradients
   and less explosive transverse expansion. The latter leads to a
   decrease of the effective temperature (inverse slope) $T_{\rm eff}$ of the
   $p_T$-spectra, see Fig.~\ref{deps1}, lower panel. Larger $R_\perp$ also results in decreased ellipticity and triangularity of an initial energy density profile, which is hydrodynamically translated into smaller final elliptic ($v_2$, see Fig.~\ref{deps2}) and triangular ($v_3$) flow components.
 \item In a similar manner, the increase of $R_\eta$ leads to shallower
   longitudinal gradients and weaker longitudinal expansion. Thus
   more energy remain at midrapidity to form stronger transverse expansion,
   which increases $T_{\rm eff}$ and $v_2$. On the other hand, larger $R_\eta$ also results in larger initial entropy of the system, which considerably increases the final particle multiplicity, see Fig.~\ref{deps1}, upper panel.
 \item Increased $\tau_0$ has two effects:
  \begin{enumerate}
  \item It leads to a shorter lifetime of the hydrodynamic phase, as a result of longer pre-thermal phase.
  \item At the same time $\tau_0$ enters the Gaussian energy/momentum smearing profile. Thus its increase acts opposite to the increase of $R_\eta$.
  \end{enumerate}
  From the response of the observables to the increase of $\tau_0$ we find that the second effect is stronger.
 \item Increased $\epsilon_{\rm sw}$ shortens the effective lifetime of the hydrodynamic
   phase. The shorter time to develop radial and elliptic flows 
   is not fully compensated by the longer cascade phase,
   which results in the decrease of both final $T_{\rm eff}$ and final $v_2$.
   Since the total entropy is conserved in the ideal hydrodynamic expansion, but increases in the cascade stage, the final particle multiplicity increases with the increase of $\epsilon_{\rm sw}$.
\end{itemize}

The observed dependencies are schematically depicted in
Table~\ref{tbDep}, where the signs of the responses of the
observables to the increase of a particular model parameter are
shown. As for the magnitudes of the response, one can see from the
plots that by varying the parameters of the initialization procedure
one has a nearly linear influence on the final $dN/dy$, $T_{\rm eff}$ and
$v_2$. From Fig.~\ref{deps2} one can see that by choosing a
larger value of $R_\perp$ it is possible to approach the experimental
value of $v_2$ with zero shear viscosity in the hydrodynamic phase. However,
such value is inconsistent with the $p_T$ spectra and charged particle
multiplicity.

\begin{table}
\begin{tabular}{|l|c|c|c|c|c|}
\hline
           ~      & $R_\perp$ $\uparrow$ & $R_z$ $\uparrow$ & $\eta/s$ $\uparrow$ & $\tau_0$ $\uparrow$ & $\epsilon_{\rm sw}$ $\uparrow$ \\ \hline
  $T_{\rm eff}$   &  $\downarrow$ & $\uparrow$ & $\uparrow$    & $\downarrow$ & $\downarrow$ \\ \hline
      $dN/dy$     &   $\uparrow$  & $\uparrow$ & $\uparrow$    & $\downarrow$ &  $\uparrow$  \\ \hline
      $v_2$       &  $\downarrow$ & $\uparrow$ & $\downarrow$  & $\downarrow$ & $\downarrow$ \\ \hline
 \end{tabular}
\caption{Schematical representation of the response (increase or decrease) of the observables to the increase of a particular parameter of the model.}\label{tbDep}
\label{table-cuts}
\end{table}

Investigating all the dependencies in detail allows us to choose
parameter values which lead to a reasonable
reproduction of the data. These values are shown in
Table~\ref{tbParameters}. For reasons of simplicity we keep
$\epsilon_{\rm sw}=0.5$~GeV/fm$^3$ for all collision energies, since
the other parameters provide enough freedom for adjustment.  Note that
since the model requires a lot of CPU time to obtain results for each
particular collision energy and centrality, it is at the moment impractical to
provide $\chi^2$-optimized values of the model parameters and their errors. Thus
the parameters are adjusted manually based on a visual correspondence
to the data. A full fledged $\chi^2$ fit to the data is planned for the future
using a model emulator, as suggested in
Refs.~\cite{Petersen:2010zt, Novak:2013bqa, Bernhard:2015hxa}.

\begin{table}
\begin{tabular}{|l|l|l|l|l|}
\hline
 $\sqrt{\sNN}$~[GeV] & $\tau_0$~[fm/c] & $R_\perp$~[fm] & $R_\eta$~[fm] & $\eta/s$ \\ \hline
     7.7          &      3.2        &     1.4        &     0.5    &    0.2   \\ \hline
     8.8 (SPS)    &      2.83       &     1.4        &     0.5    &    0.2   \\ \hline
     11.5         &      2.1        &     1.4        &     0.5    &    0.2   \\ \hline
     17.3 (SPS)   &      1.42       &     1.4        &     0.5    &    0.15  \\ \hline
     19.6         &      1.22       &     1.4        &     0.5    &    0.15  \\ \hline
     27           &      1.0        &     1.2        &     0.5    &    0.12  \\ \hline
     39           &      0.9*        &     1.0        &     0.7    &    0.08  \\ \hline
     62.4         &      0.7*        &     1.0        &     0.7    &    0.08  \\ \hline
     200          &      0.4*        &     1.0        &     1.0    &    0.08  \\ \hline
 \end{tabular}
\caption{Collision energy dependence of the model parameters chosen to
  reproduce the experimental data in the BES region and higher RHIC
  energies. An asterisk denotes the values of starting time $\tau_0$ which are adjusted instead of being taken directly from Eq.~\ref{eqTau0}.}\label{tbParameters}
\end{table}

\section{Results for Bulk Observables}
\label{results}

\begin{figure}[!htb]
\includegraphics[width=0.47\textwidth]{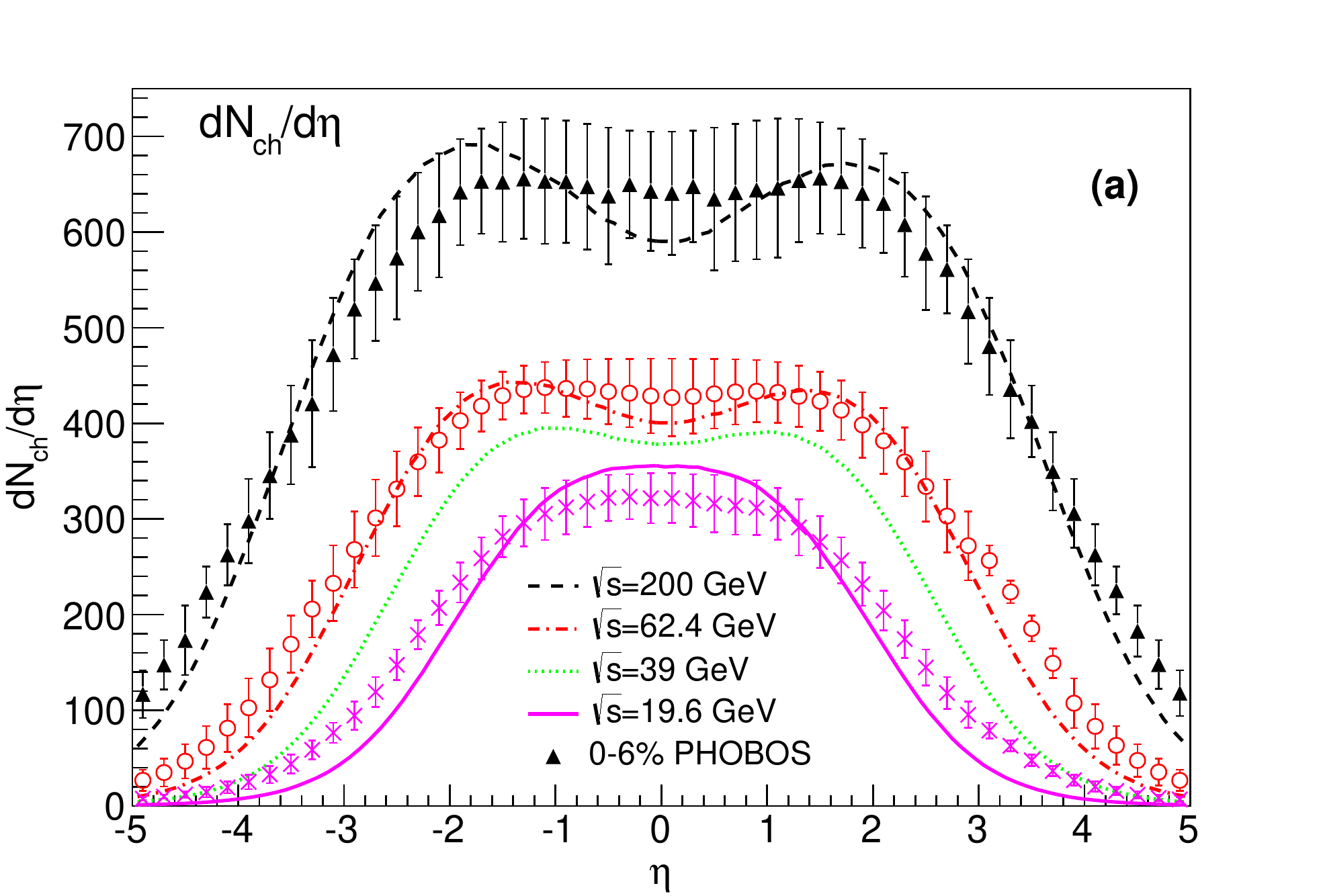}\\
\includegraphics[width=0.47\textwidth]{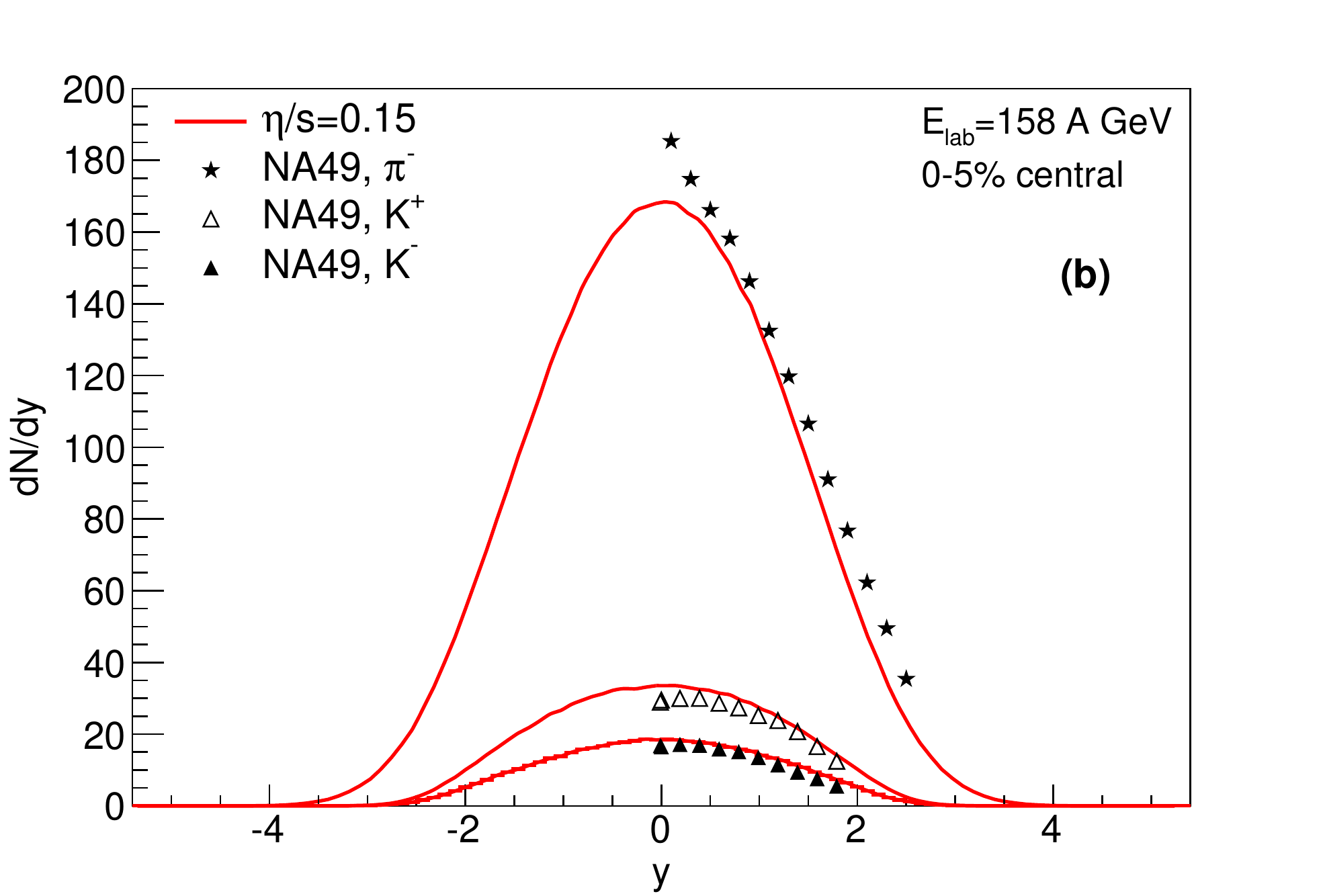}\\
\includegraphics[width=0.47\textwidth]{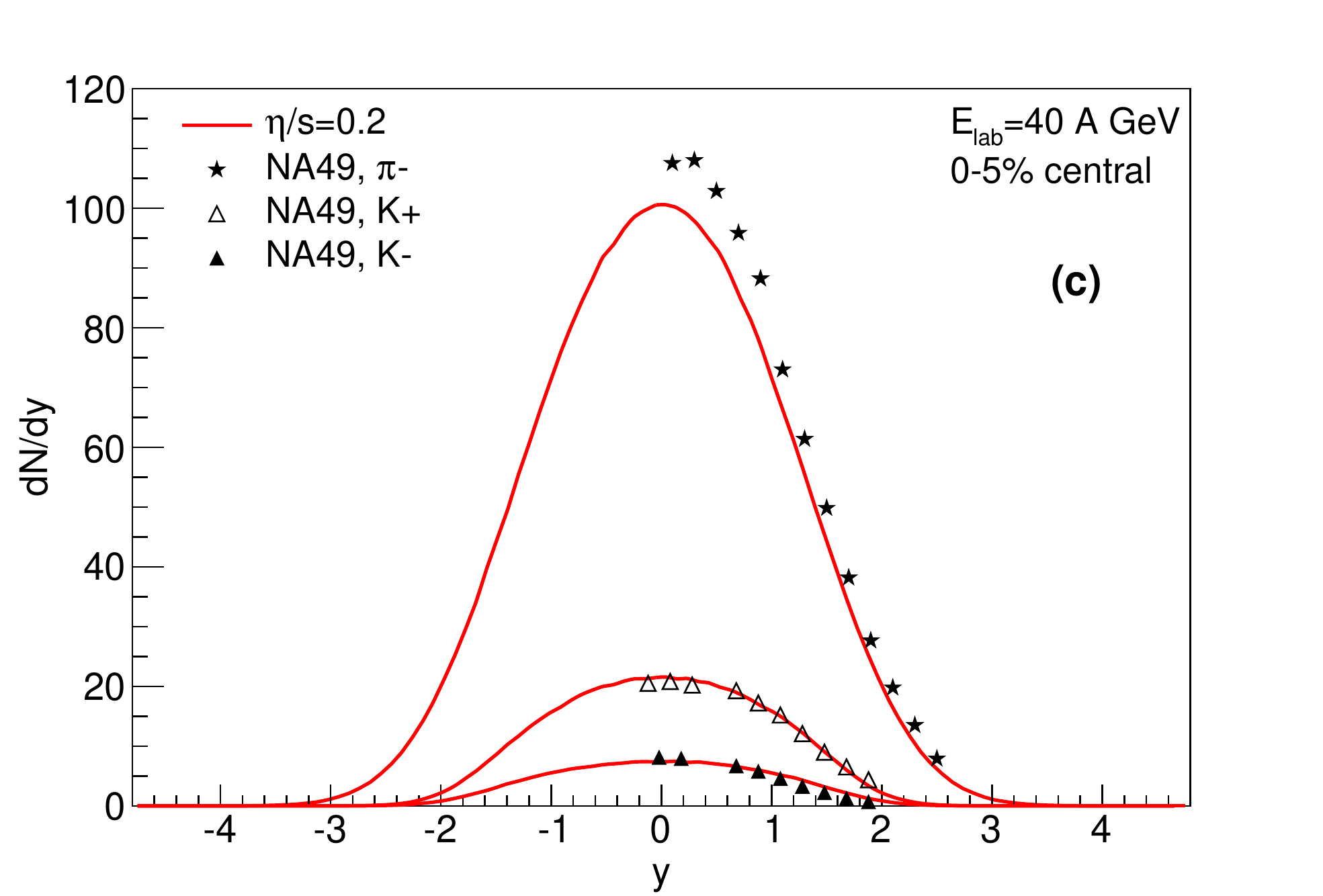}
\caption{(Color online) Pseudorapidity distributions of charged hadrons (top) in
  Au-Au collisions at $\sqrt{\sNN}=19.6$, 39,
  62.4 and 200~GeV energies, and rapidity distributions of
  identified hadrons in Pb-Pb collisions at
  $E_{\rm lab}=158$ and 40~A~GeV ($\sqrt{\sNN}=17.6$ and 8.8 GeV)
  energies (middle and bottom panels, respectively). The calculations were done
  using the collision energy dependent parameters listed in
  Table~\ref{tbParameters}. The data are from the
  PHOBOS~\cite{Alver:2010ck} and the NA49~\cite{Afanasiev:2002mx}
  collaborations.}\label{figDndyFit}
\end{figure}

\begin{figure}[!htb]
\includegraphics[width=0.47\textwidth]{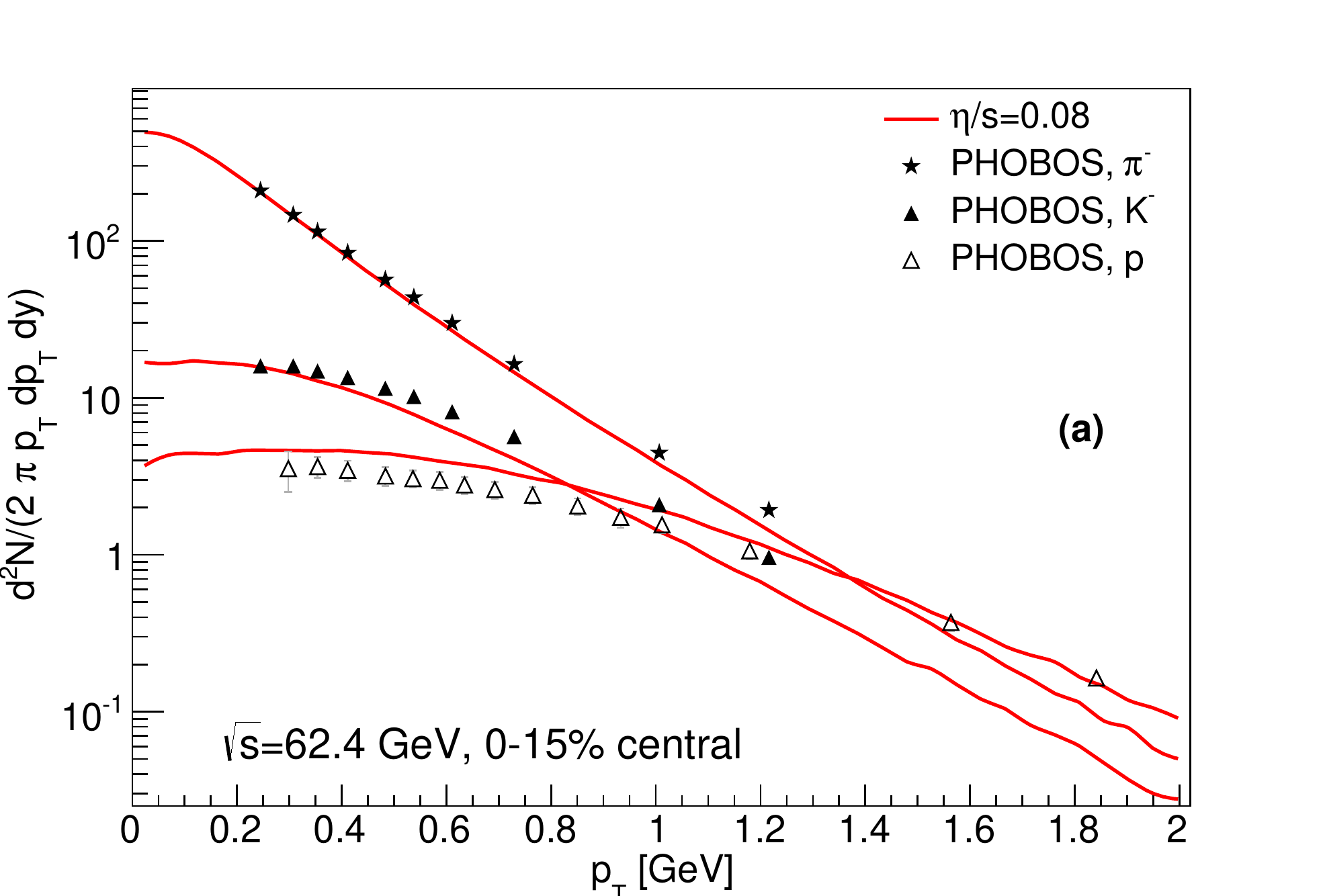}\\
\includegraphics[width=0.47\textwidth]{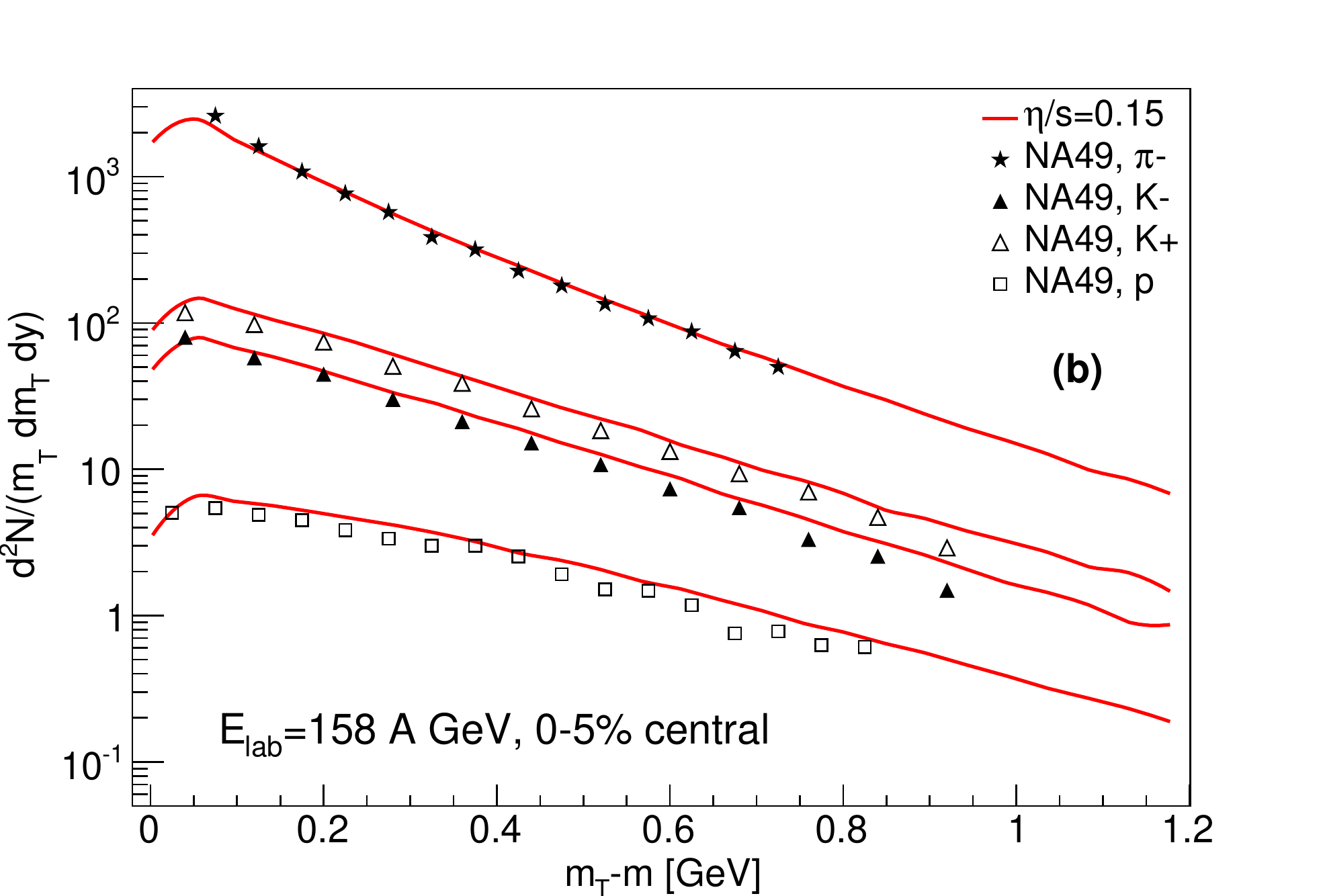}\\
\includegraphics[width=0.47\textwidth]{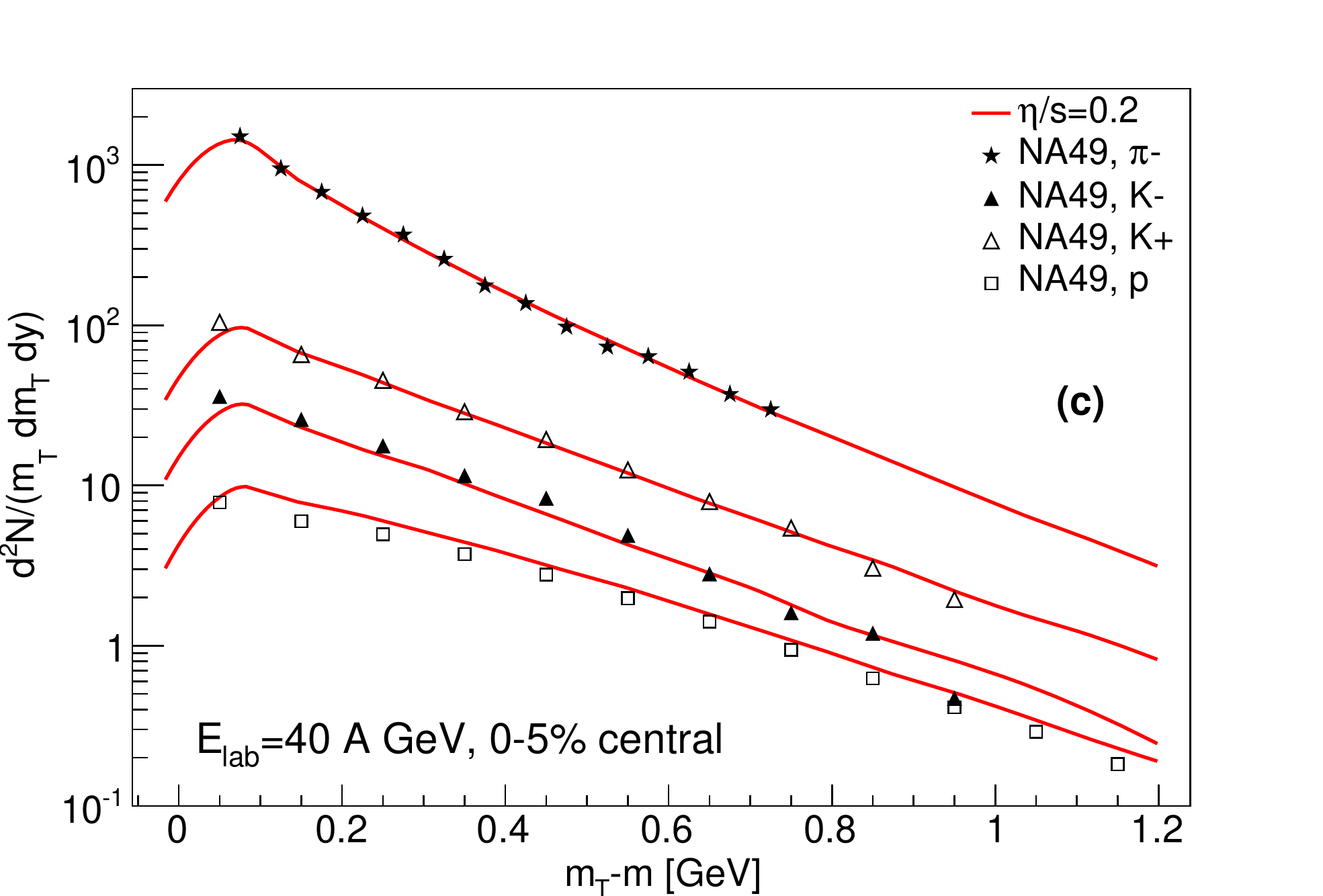}
\caption{(Color online) $p_T$ spectra of identified hadrons in Au-Au collisions 
at $\sqrt{\sNN}=62.4$ GeV energy (top) and in Pb-Pb collisions at
  $E_{\rm lab}=158$ and 40~A~GeV ($\sqrt{\sNN}=17.6$ and 8.8 GeV)
  energies (middle and bottom panels, respectively). The model
  calculations were carried out using the collision energy dependent
  parameters listed in Table~\ref{tbParameters}, and the data are from
  the PHOBOS and NA49
  collaborations~\cite{Back:2006tt,Afanasiev:2002mx,Anticic:2010mp}.}\label{figPtFit}
\end{figure}

\begin{figure}
\includegraphics[width=0.47\textwidth]{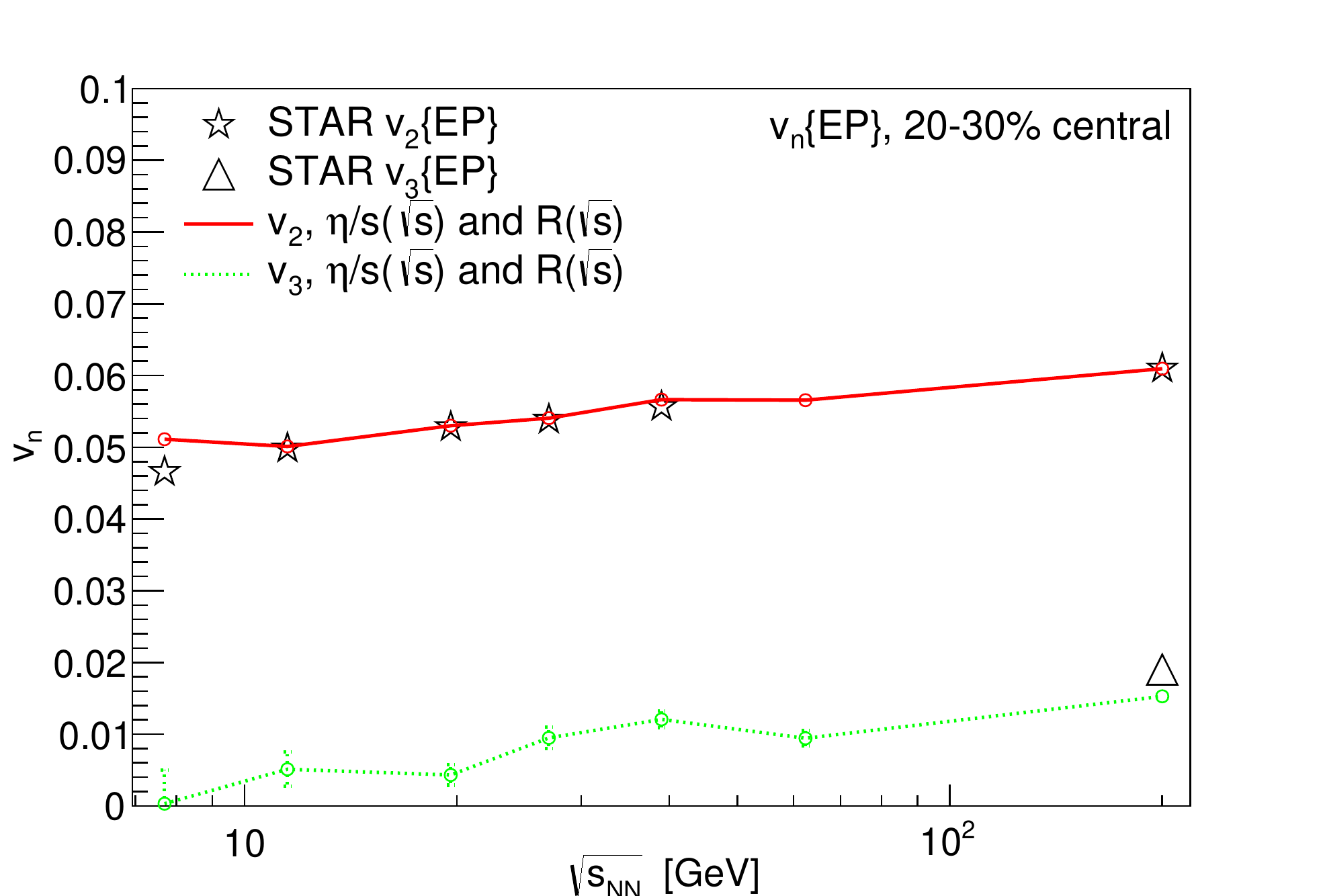}
\caption{(Color online) $p_T$ integrated elliptic and triangular flow 
coefficients $v_2$ and $v_3$ as a function of collision energy. Both the
  experimental and calculated coefficients were evaluated using the
  event plane method. The calculation was done using the collision
  energy dependent parameters listed in Table~\ref{tbParameters}, and
  the data is from the STAR collaboration~\cite{Adamczyk:2012ku,Adamczyk:2013waa}.}
\label{figv2Fit}
\end{figure}

Finally, let us have a look at the results for bulk observables with
the energy dependent parameters for the hydrodynamic description (see
Table~\ref{tbParameters}).

The (pseudo)rapidity spectra are presented in
Fig.~\ref{figDndyFit}. One can see that whereas the parameters were
adjusted to reproduce the total multiplicities, the resulting shapes
of the pseudorapidity distributions are also in a reasonable agreement
with the data. From the model results one can observe the change in
shape from the single peak structure at $\sqrt{\sNN}<20$~GeV to
a doubly-peaked distribution (or from a Dromedary to a Bactrian camel shape) 
which starts to form 
at $\sqrt{\sNN}=39$~GeV.  At higher collision energies we observe
a shallow dip around zero pseudorapidity.

The $p_T$ spectra of pions, kaons and protons in collisions at
$\sqrt{\sNN}=62.4$, 17.6, and 8.8 GeV energies are shown in
Fig.~\ref{figPtFit}. In general the spectra and especially the $p_T$
slopes are reproduced, which indicates that both the collective radial
flow (generated in the hydrodynamic and cascade stages), and thermal motion are
combined in the right proportion.

\begin{figure}
\includegraphics[width=0.47\textwidth]{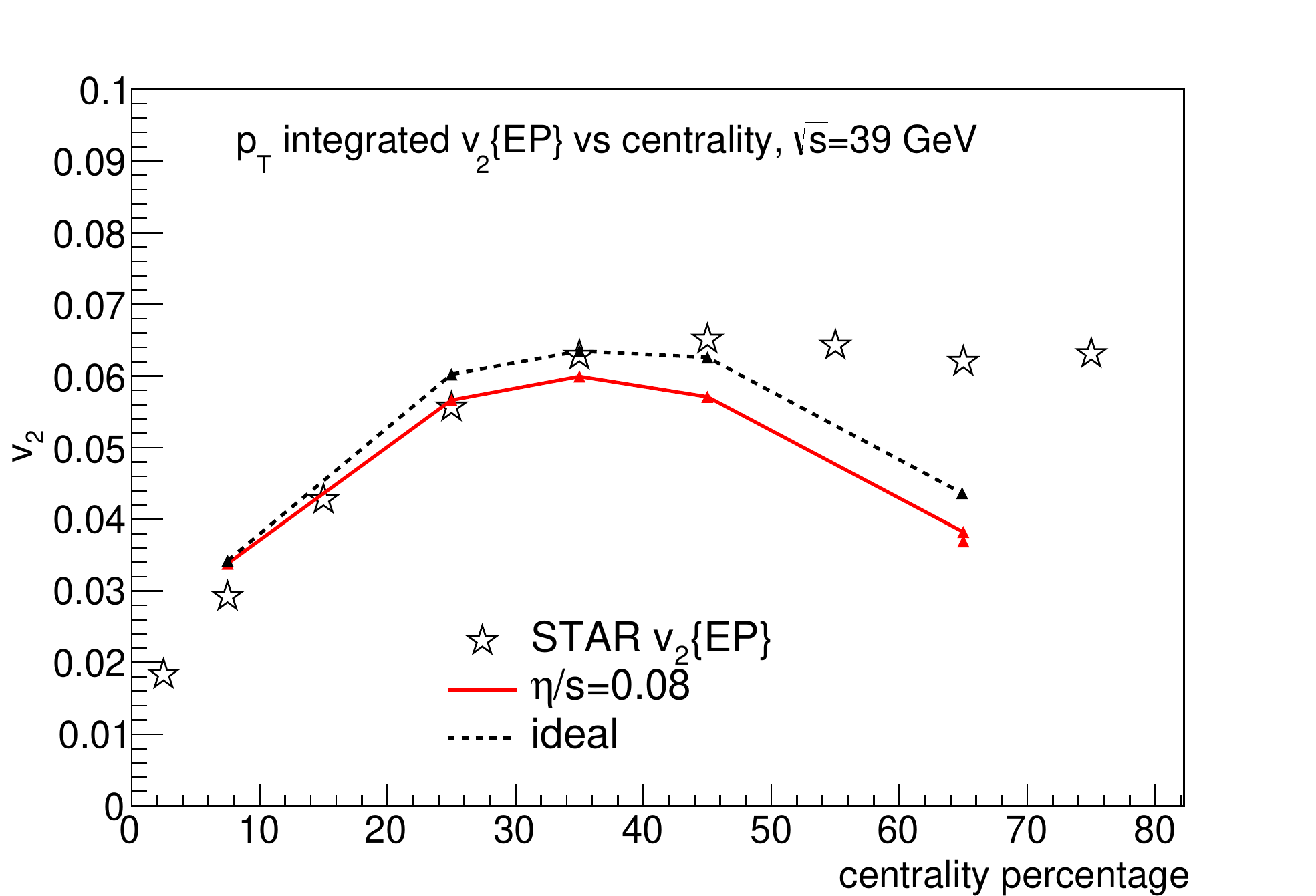}
\caption{(Color online) $p_T$ integrated elliptic flow coefficient $v_2$ in
  $\sqrt{\sNN}=39$~GeV Au-Au collisions as function of
  centrality. Both the experimental and calculated $v_2$ was evaluated
  using the event plane method. The calculation was done using the
  collision energy dependent parameters listed in
  Table~\ref{tbParameters}, and the data is from the STAR
  collaboration~\cite{Adamczyk:2012ku}.}\label{figv2rom}
\end{figure}

The elliptic and triangular flow coefficients for 20-30\% central
Au-Au collisions as a function of collision energy are presented in
Fig.~\ref{figv2Fit}. As expected, the calculated values of the elliptic flow
follow the data closely,
since this quantity was used to fix the parameters. In contrast to that,
triangular flow $v_3$ is calculated from the same simulated
events, and thus can be considered as a prediction of the
model. We expect that the non-monotonous behavior of $v_3$ is
an artifact of our fitting procedure, and more careful adjustment of the model parameters would further smoothen the behavior of $v_3(\sqrt{s})$.

\begin{figure}
\includegraphics[width=0.47\textwidth]{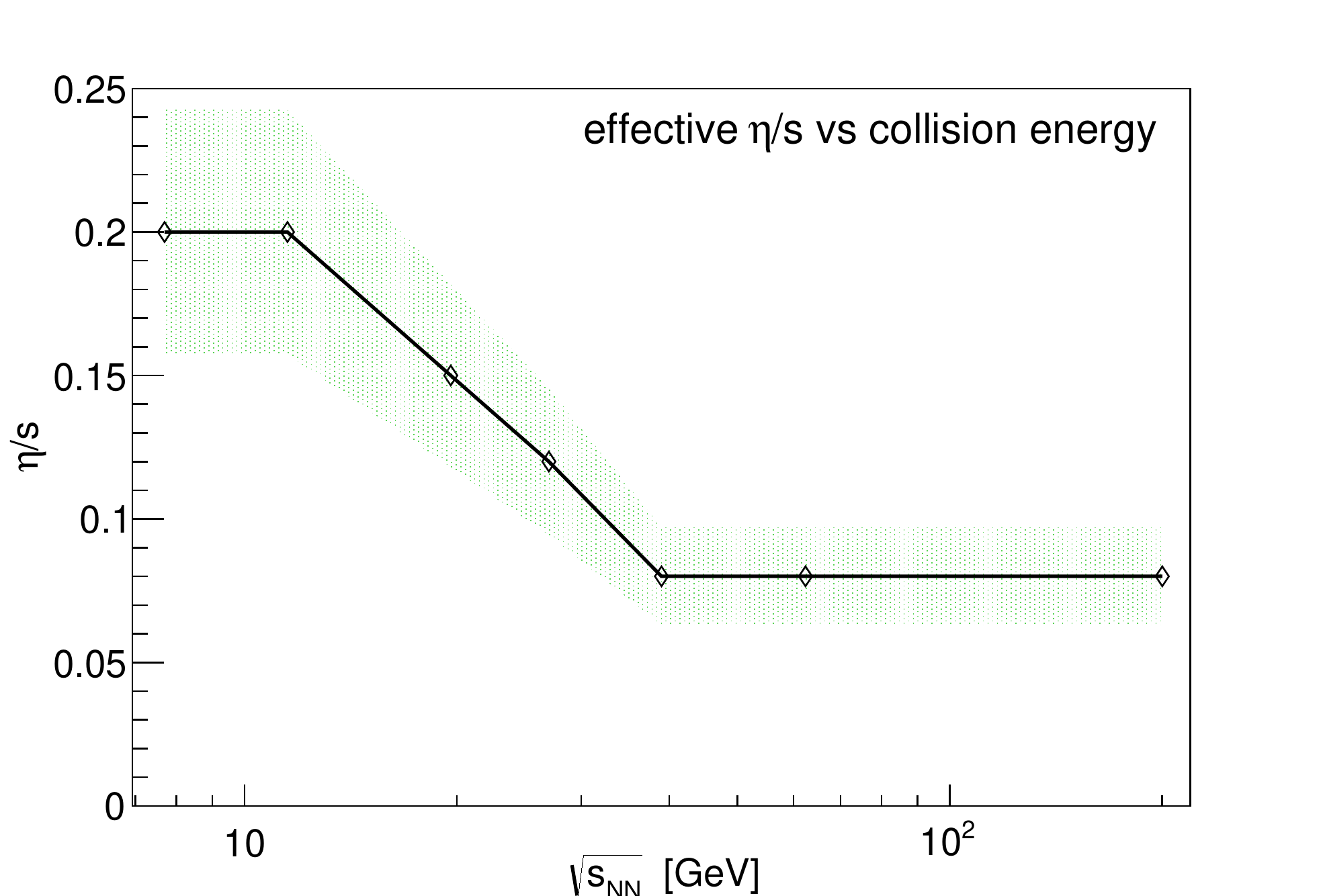}
\caption{(Color online) Effective values of shear viscosity over entropy density 
$\eta/s$ used to describe the experimental data at different collision energies 
as shown in Table~\ref{tbParameters}. The green band represent an estimate of 
uncertainty in $\eta/s$ resulting from the allowed variation of model parameters 
around their optimal values.}
\label{figEtaS}
\end{figure}

The 20-30\% centrality class was chosen because the elliptic flow signal is strongest around this centrality class. Also, at this centrality nonflow contributions from minijets, which are not included in the model, are small.
 The centrality dependence of elliptic flow at $\sqrt{\sNN}=39$~GeV is shown in
Fig.~\ref{figv2rom}. The parameters are the same at all
centralities. In peripheral collisions the model significantly undershoots
the data. This is due to the smoothening procedure used
to convert individual particles to the fluid-dynamical initial
state. With the present smearing parameters
the eccentricity of the system is too small in peripheral
collisions, where the size of the entire system is comparable
to the smearing radius.

The most important conclusion from the adjustment procedure is that
reproduction of the data requires an effective $\eta/s$ which
decreases as a function of increasing collision energy, see Table~\ref{tbParameters} and Fig.~\ref{figEtaS}. On Fig.~\ref{figEtaS} one can also see an estimated error band around the optimal values of $\eta/s$. As mentioned, a proper determination of the error bars would require a $\chi^2$ fit. Currently the error band is estimated from the variations of two parameters of the model ($\eta/s$ and $R_T$) which result in the same value of $p_T$ integrated elliptic flow and a 5\% variation in the slope of proton $p_T$ spectrum, which is the most sensitive to a change in radial flow.

In the present calculations $\eta/s$ is taken to be constant during the
evolution of the system, and its value changes only with the collision
energy. However we expect that physical $\eta/s$ depends on both the temperature and baryon chemical potential,
and that $\eta/s$ has a minimum around $T_c$ and zero $\mu_b$ 
\cite{Csernai:2006zz, Denicol:2010tr, Niemi:2011ix, Niemi:2012ry}. The smaller 
the collision energy, the larger the average baryon chemical potential in the 
system. This indicates that the physical value of $\eta/s$ should increase with 
increasing $\mu_B$.

\begin{figure}
 \includegraphics[width=0.47\textwidth]{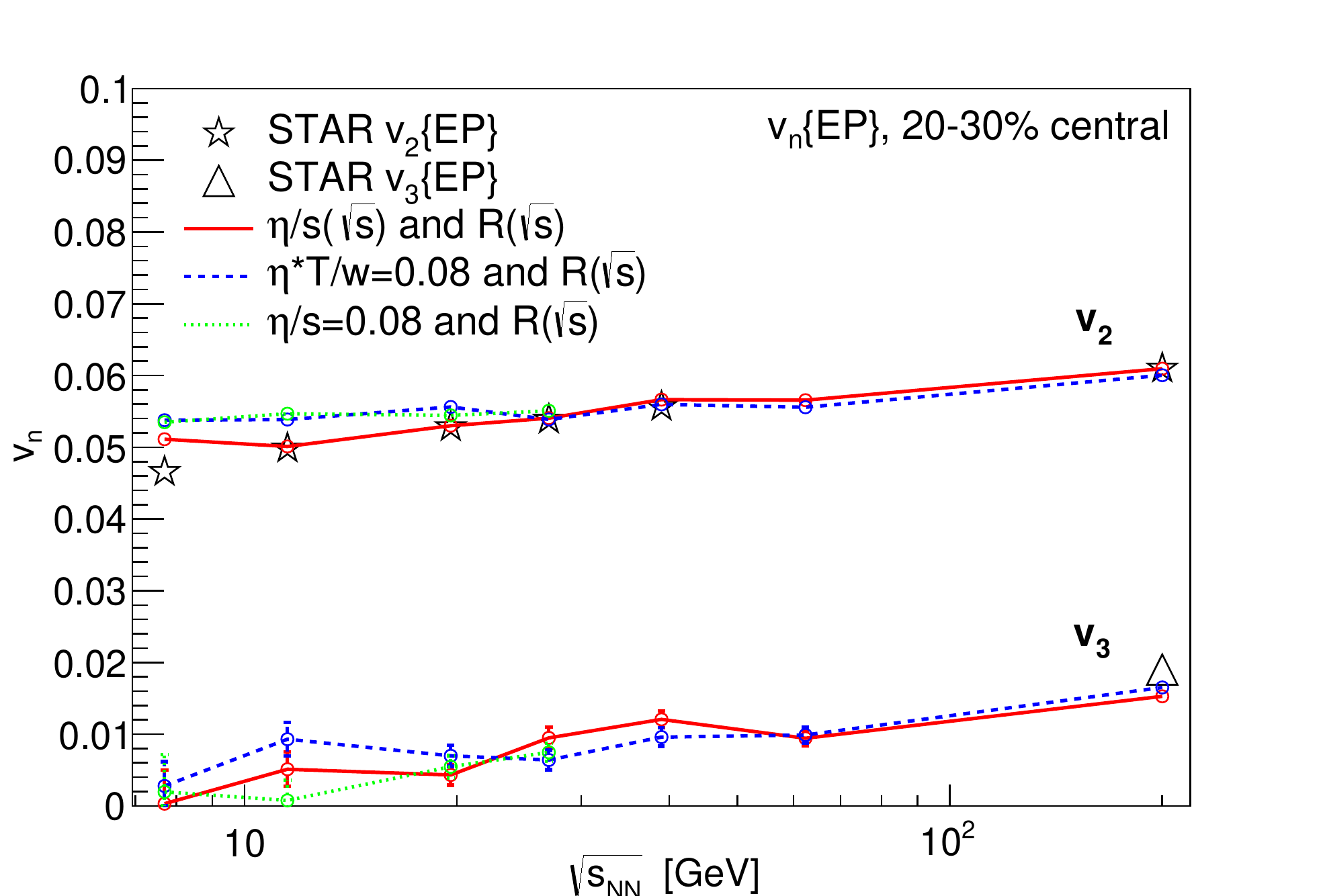}
 \caption{(Color online) $p_T$ integrated elliptic and triangular flow 
coefficients $v_2$ and $v_3$ as a function of collision energy. Solid red line
   represents the results from Fig.~\ref{figv2Fit} obtained using
   collision energy dependent $\eta/s$. Dashed blue and dotted green
   lines correspond to collision independent $\eta T/w=0.08$ and
   $\eta/s=0.08$, respectively. In all three cases the other model
   parameters were taken to depend on the collision energy as shown
   in Table~\ref{tbParameters}. The experimental data is from the STAR
   collaboration~\cite{Adamczyk:2012ku,Adamczyk:2013waa}.}
\label{figv2v3_etaW}
\end{figure}

In Ref.~\cite{Liao:2009gb} it was argued that $\eta/s$ is not an
appropriate measure of the fluidity of the system. However, the measure of
fluidity proposed in that paper, $L_\eta/L_n=(\eta n^{1/3})/(w c_s)$, 
where $n$ is the total particle number density, $w$ enthalpy, and
$c_s$ the speed of sound, is difficult to implement in the present
fluid-dynamical calculation since $n$ is not well defined in our
two-phase EoS. Instead, we use as an alternative measure of fluidity
the combination $\eta T/w = \eta T/(\epsilon + P) =
\eta/(s+\sum_\alpha \mu_\alpha n_\alpha/T)$, where $n_\alpha$ are
the charge densities (baryon, strange, electric) and $\mu_\alpha$ the 
corresponding chemical potentials, and which approaches $\eta/s$ in
the limit of small charge densities. We have performed an additional
round of simulations, keeping $\eta T/w=0.08$ and $\eta/s=0.08$ at all
collision energies to see whether different measure of fluidity makes
any difference. The resulting elliptic and triangular flow
coefficients are shown in Fig.~\ref{figv2v3_etaW}. One can see that at
all considered collision energies there is no visible difference in
the elliptic flow coefficient between the $\eta/s=0.08$ and $\eta
T/w=0.08$ cases. We have also checked that the two scenarios result in
virtually same $p_T$ spectra and $dN/dy$ distributions. This indicates
that the contribution from baryon/electric charge density to the
entropy density does not induce strong enough baryon density
dependence of the $\eta/s$ ratio to affect the hydrodynamic evolution.

\section{Summary and Outlook}
\label{summary}

A hybrid model featuring a 3+1-dimensional viscous hydrodynamic
phase with an explicit treatment of finite baryon and charge
densities is introduced. The model employs a chiral model equation of state
for the hydrodynamic stage. The initial and late non-equilibrium stages are
modeled using the UrQMD hadron cascade on an event-by-event basis.

This hybrid model was applied to describe the dynamics of relativistic
heavy ion collisions at energies ranging from the lowest RHIC beam
energy scan energy to full RHIC energy, $\sqrt{s}=7.7-200$~GeV.
After tuning the
parameters, it was possible to reproduce the observed pseudorapidity
and transverse momentum distributions of produced hadrons and their
elliptic flow coefficients. The reproduction of the data requires
a finite shear viscosity over entropy density ratio $\eta/s$ which
depends on collision energy. This ratio was found to decrease from
$\eta/s=0.2$ to $0.08$ as collision energy increases from $\sqrt{\sNN} = 7.7$
to $39$~GeV, and to stay at $\eta/s=0.08$ for $39\le\sqrt{s}\le200$~GeV. 
Since the average baryochemical potential at midrapidity
decreases with increasing collision energy, the required collision
energy dependence of the effective $\eta/s$ indicates that the
physical $\eta/s$-ratio may depend on baryochemical potential, and that
$\eta/s$ increases with increasing $\mu_B$. It was also found that a
constant and collision energy independent $\eta T/w=0.08$ and $\eta/s=0.08$
in hydrodynamic phase yield quantitatively similar results. This
indicates that the $\mu_{\rm B}n_{\rm B}$ term in entropy density does
not induce the baryon density dependence of $\eta/s$ required to
reproduce the data when $\eta T/w$ is kept independent of collision
energy.

In addition we have explored the parameter dependence of the model results
and generally found a $<10$\% variation of the results, when the individual
parameters were varied by 10\%.
Of course, the proper evaluation of the effect
of finite baryochemical potential on $\eta/s$ would require
reproducing all the data using the same temperature and baryochemical
potential dependent parametrization of $\eta/s$ at all energies and
centralities. This will be addressed in future studies.

\begin{acknowledgments}
  The authors acknowledge the financial support by the Helmholtz
  International Center for FAIR and Hessian LOEWE initiative. The work
  of P.H.\ was supported by BMBF under contract no.~06FY9092. H.P.\
  acknowledges funding by the Helmholtz Young Investigator Group
  VH-NG-822 from the Helmholtz Association and GSI. Computational resources have been provided by the Center
  for Scientific Computing (CSC) at the Goethe University Frankfurt.
\end{acknowledgments}

\end{document}